\documentclass[aps,prl,superscriptaddress,twocolumn,citeautoscript,amsmath, amssymb, longbibliography]{revtex4-1}

\usepackage{graphicx}
\usepackage{dcolumn}
\usepackage{bm}
\usepackage{csquotes}
\usepackage[usenames, dvipsnames]{color}
\usepackage{subcaption}
\usepackage{tikz}
\usetikzlibrary{fit}
\usepackage{caption}
\usepackage{balance}
\usepackage[percent]{overpic}
\usepackage[colorlinks=true,
            linkcolor=blue,
            citecolor=blue,
            urlcolor=blue
]{hyperref}
\usepackage{xr-hyper}
\def\vR{\mathbf{R}}
\def\vq{\mathbf{q}}

\begin{document}

\preprint{APS/123-QED}

\title{Spin Hamiltonian, magnetic anisotropy and excitation spectrum in \\ altermagnetic $\alpha$-MnTe from first principles}

\author{Matteo Costa}
\address{CPHT, CNRS, École polytechnique, Institut Polytechnique de Paris, 91120 Palaiseau, France}
\address{Collège de France, Université PSL, 11 place Marcelin Berthelot, 75005 Paris, France}

\author{Dario Fiore Mosca}
\address{University of Vienna, Faculty of Physics, Center for Computational Materials Science, Kolingasse 14-16, 1090, Vienna, Austria}

\author{Cesare Franchini}
\address{University of Vienna, Faculty of Physics, Center for Computational Materials Science, Kolingasse 14-16, 1090, Vienna, Austria}
\address{Department of Physics and Astronomy \enquote{Augusto Righi}, Alma Mater Studiorum - Università di Bologna, Bologna, 40127 Italy}

\author{Leonid V. Pourovskii}
\address{CPHT, CNRS, École polytechnique, Institut Polytechnique de Paris, 91120 Palaiseau, France}
\address{Collège de France, Université PSL, 11 place Marcelin Berthelot, 75005 Paris, France}

\date{\today}

\begin{abstract}
The hexagonal altermagnet $\alpha$-MnTe combines a near-room-temperature collinear order with a strongly spin-split electronic structure, yet the microscopic origin of its easy-plane anisotropy, its zone-center magnon gap and its anomalously large magnetostriction have remained unexplained. We derive the complete low-energy magnetic Hamiltonian of $\alpha$-MnTe from first principles, combining density functional theory and dynamical mean-field theory with a force-theorem evaluation of the full anisotropic intersite exchange, single-ion terms and classical dipolar couplings. The isotropic exchange, including the altermagnetic splitting of the tenth-neighbor coupling, agrees with inelastic neutron scattering data. Contrary to the common single-ion picture, we find that the easy-plane anisotropy and the magnon gap at $\Gamma$ are governed by classical dipole-dipole interactions, which enhance the zone-center magnon splitting threefold leading to quantitative agreement with experiment, while the Dzyaloshinskii-Moriya term is irrelevant despite its size. The same Hamiltonian reproduces the chiral magnon splitting along antinodal directions, mirroring the electronic spin splitting, and quantitatively accounts for the giant magnetovolume coupling of $\alpha$-MnTe through the volume dependence of superexchange.
\end{abstract}

\maketitle

\paragraph*{Introduction.} After decades of binary classification of collinear magnetic systems as ferromagnets or antiferromagnets, the scientific community has recently come to realize that there exists a third, qualitatively different, type of magnetic order. This third class of collinear magnetic systems is now the subject of growing research interest~\cite{bai2024,song2025,jungwirth2025,jungwirth2026,yan2024,tamang2025} and it is widely known as $"$altermagnetism$"$ inspired by the characteristic alternating spin-splitting of the electronic bands across the first Brillouin zone. Not only this momentum dependence constitutes the defining feature of altermagnets, but it also implies that the breaking of time-reversal symmetry coexists with the zero net magnetization: two properties that were long considered mutually exclusive in collinear magnetic systems. Although several theoretical works had previously predicted time-reversal-symmetry-breaking phenomena~\cite{smejkal2022hall,smejkal2020,naka2019,feng2022,reichlova,mazin2021,gonzales2021,naka2021,bose2022,smejkal2022giant,shao2021,ma2021} and spin-split electronic bandstructures~\cite{noda2016,okugawa2018,ahn2019,hayami2019,yuan2020,hayami2020,yuan2021,egorov2021,mazin2021,naka2021,reichlova,ma2021,gonzales2021,shao2021} in unconventional antiferromagnets, it was only through recent seminal studies~\cite{smejkal2022,smejkal2022review,krempasky2024} that a unified description of the three collinear magnetic phases could be established within the framework of nonrelativistic spin-group theory.

\begin{figure*}[ht]
\centering
\begin{minipage}[c]{0.29\textwidth}
\centering
\begin{subfigure}{\linewidth}
\centering
\begin{tikzpicture}
    \node[anchor=south west,inner sep=0] (img) at (1.0,0)
        {\includegraphics[width=\linewidth]{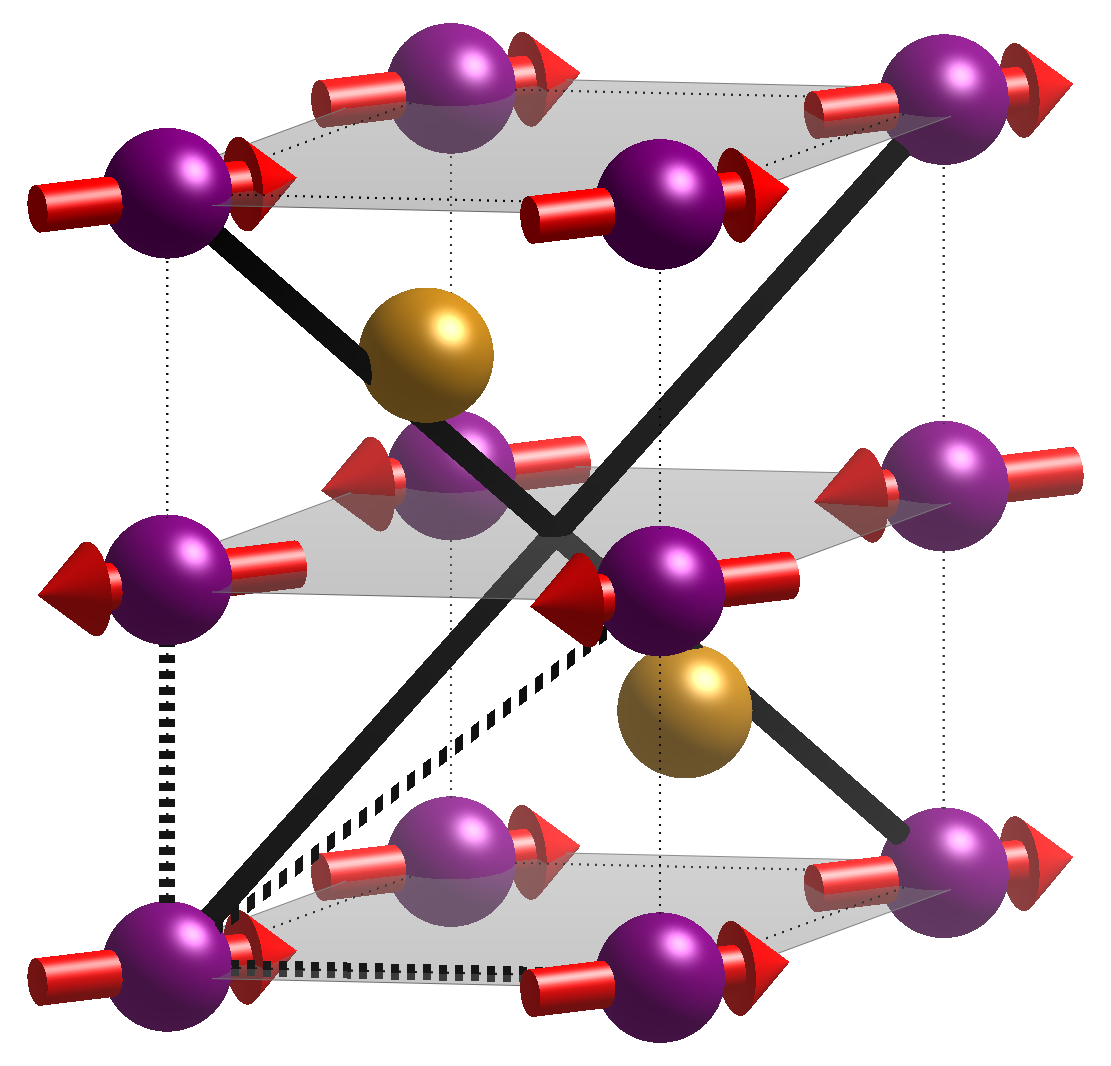}};
    \begin{scope}[x={(img.south east)},y={(img.north west)}]
        \node[black] at (0.20,0.3) {${\text{J}}_1$};
        \node[black] at (0.48,0.03) {${\text{J}}_2$};
        \node[black] at (0.53,0.3) {${\text{J}}_3$};
        \node[black] at (0.33,0.32) {${\text{J}}_{10}$};
        \node[black] at (0.75,0.35) {${\text{J'}}_{10}$};
        \node[white,font=\bfseries] at (0.272,0.093) {Mn};
        \node[white,font=\bfseries] at (0.375,0.665) {Te};
    \end{scope}
\end{tikzpicture}
\begin{tikzpicture}
\begin{scope}[local bounding box = paneltot]
\begin{scope}[
	    shift={(0,0)},
	    x={(-0.5*0.6cm,-0.866*0.6cm)},
	    y={(1cm,0cm)},
	    z={(0cm,1cm)},
	    >=stealth,
	    scale=0.9
	]
	
	\def\thet{45}
	\def\L{1.5}
	\def\R{0.7}
	
	\coordinate (O) at (0,0,0);
	\coordinate (A) at ({cos(30)},{-sin(30)},0);
	\coordinate (B) at (0,1,0);
	\coordinate (AB) at ({cos(30)},{1-sin(30)},0);
	
	\coordinate (C) at (0,0,1);
	\coordinate (AC) at ({cos(30)},{-sin(30)},1);
	\coordinate (BC) at (0,1,1);
	\coordinate (ABC) at ({cos(30)},{1-sin(30)},1);
	
	\draw[gray, dashed, thin] ({2*sin(\thet)},0,{2*cos(\thet)})--(0,0,{2*cos(\thet)});
	\draw[gray, dashed, thin] ({2*sin(\thet)},0,{2*cos(\thet)})--({2*sin(\thet)},0,0);
	
	\draw[very thick,->] (O)--(A)
	    node[below left] {$\mathbf a$};
	
	\draw[dashed, very thick, ->] (O)--(1.7,0,0)
	    node[below left] {$x$};
	
	\draw[->,very thick] (O)--(0,1.45,0)
	    node[below left] {$\mathbf{b} \parallel y$};
	
	\draw[->,very thick] (O)--(0,0,1.65)
	    node[above] {$\mathbf{c} \parallel z$};
	
	\draw[very thick, red, ->]
	    (O)--({2*sin(\thet)},0,{2*cos(\thet)});
	\node[red, above left] at ({2*sin(\thet)},0.15,{2*cos(\thet)}) {$\hat{\mathbf{m}}(\theta)$};
	
	\draw[red, thick]
	    plot[domain=0:\thet,samples=40,variable=\t]
	    ({\R*sin(\t)},0,{\R*cos(\t)});
	\node[red, below right] at (0,0,{\R}) {$\theta$};

\end{scope}
\begin{scope}[
        shift={(2.3,0)},
        scale=1,
        level/.style={thick},
        label/.style={font=\small}
    ]

    \node[label, left, align=center] at (0.95,1.57) {$^6S_{5/2}$ \\ sextet};
    \node[label, left, align=center] at (1.5,1.57) {$M_J$\\$\downarrow$};
    
    \foreach \y/\name in {
        3.0*0.6/{\pm\frac{5}{2}},
        3.0*0.2/{\pm\frac{3}{2}},
        3.0*0.0/{\pm\frac{1}{2}}
    } {
        \draw[level, very thick] (0.0,\y-0.9+0.02) -- (0.37,\y-0.9+0.02);
        \draw[level, very thick] (0.43,\y-0.9+0.02) -- (0.8,\y-0.9+0.02);
        \node[label, right] at (0.8,\y-0.9+0.02) {$\name$};
    }
    \draw[dashed, gray] (-0.85,-1)--(-0.85,1.8);
    \draw[dashed, <->] (0.6,-0.86+0.02)--(0.6,-0.94+3.0*0.2+0.02)
        node[midway, left] {\small $20$ $\mu$eV};
    \draw[dashed, <->] (0.6,-0.86+3.0*0.2+0.02)--(0.6,-0.94+3.0*0.6+0.02)
        node[midway, left] {\small $40$ $\mu$eV};

\end{scope}
\node[draw, rectangle, rounded corners=2pt, line width=0.5pt, fit=(paneltot), inner sep=2pt] {};
\end{scope}
\end{tikzpicture}
\end{subfigure}
\put(-142,230){(a)}
\put(-142,85){(b)}
\put(-63,85){(c)}
\end{minipage}
\begin{minipage}[c]{0.29\textwidth}
    \centering
    \begin{overpic}[width=\linewidth]{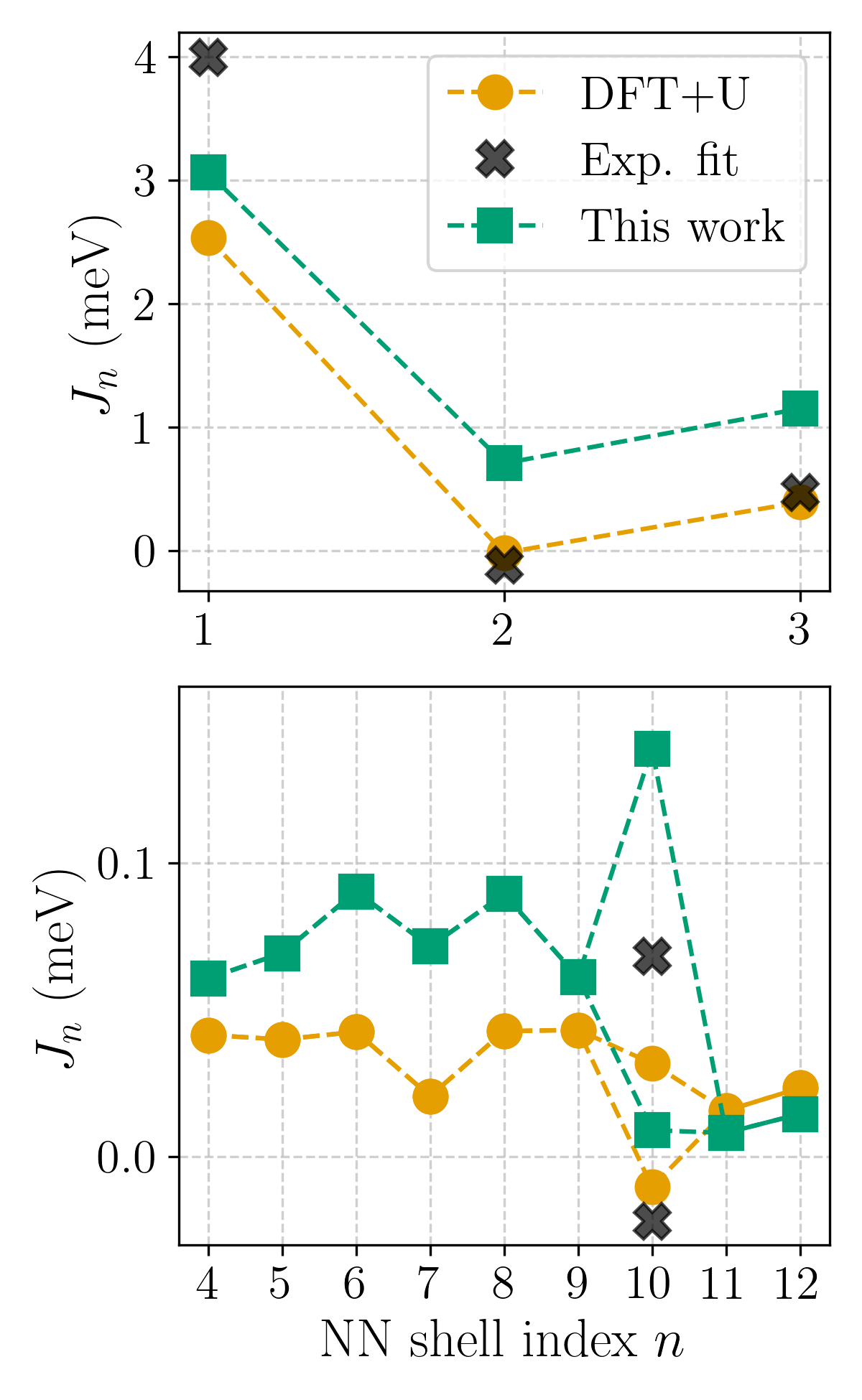}
    \put(0,95){(d)}
    \put(0,48){(e)}
    \end{overpic}
\end{minipage}
\begin{minipage}[c]{0.365\textwidth}
    \centering
    \begin{subfigure}{\linewidth}
        \begin{overpic}[width=\linewidth]{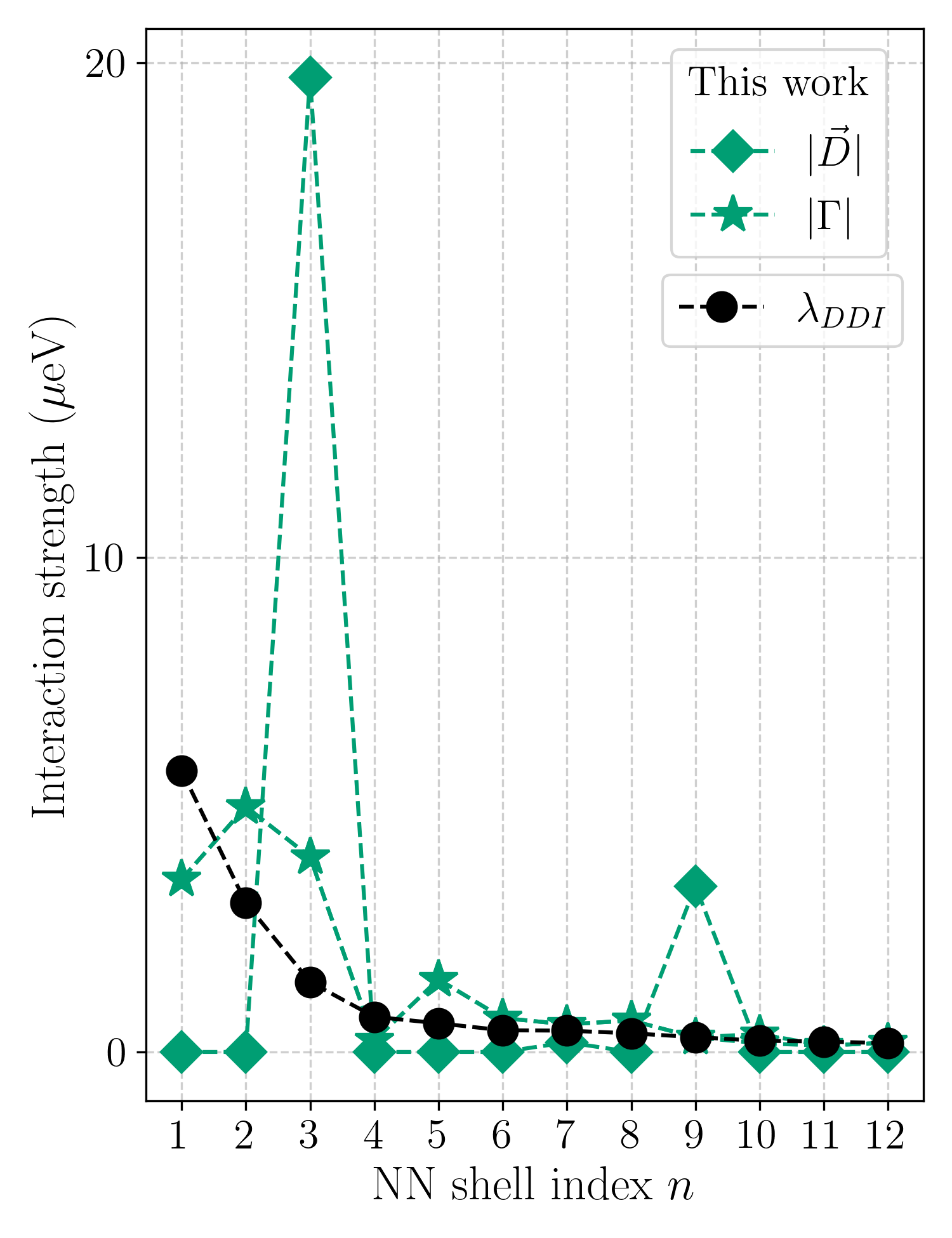}
        \put(0,95){(f)}
        \end{overpic}
    \end{subfigure}
\end{minipage}

\caption{(a) Hexagonal unit cell of MnTe, showing the first three and tenth nearest-neighbor (NN) bonds (dashed and solid lines) as well as the A-type layered antiferromagnetic order (red arrows). (b) Lattice vectors and Cartesian axes directions adopted; the azimuthal angle $\theta$ in the $ac$ plane is also indicated. (c) Calculated splitting of the Mn $^6S_{5/2}$ ground state multiplet. (d-e) Calculated isotropic contributions to the intersite exchange interactions (IEI) as a function of the NN shell index $n$, compared with the corresponding parameters extracted from inelastic neutron-scattering measurements~\cite{liu2024} and DFT+U calculations~\cite{alaei2025}: (d) for the first three NN shells, (e) for distant NN shells. (f) Magnitude of the symmetric and antisymmetric anisotropic IEI as a function of the NN shell index $n$, compared with the prefactor $\lambda_{DDI}=\frac{g^2\mu_0\mu_B^2}{4\pi|R^{(n)}|^3}$ of the classical dipole–dipole interaction (DDI).}
\label{fig:iei}
\end{figure*}

Among the many altermagnetic candidates currently under investigation, we focus on the hexagonal phase of MnTe ($\alpha$-MnTe), a near-room-temperature altermagnetic Mott insulator that crystallizes in a NiAs-type hexagonal crystal structure (space group: $P6_3/mmc$)~\cite{furberg1953,kunitomi1964,kriegner2017} at ambient conditions. Below 307 K $\alpha$-MnTe exhibits an A-type antiferromagnetic (AFM) order, consisting of ferromagnetically-ordered Mn layers alternating along the $c$-direction with an easy-plane anisotropy for the local magnetic moments (see Figs.~\ref{fig:iei}a-b)~\cite{kunitomi1964,kriegner2017}. A remarkably large altermagnetic spin splitting in the $\alpha$-MnTe electronic band structure, which is only exceeded by CrSb and RuO$_2$~\cite{smejkal2022review, ding2024}, has attracted considerable interest to this system. Accordingly $\alpha$-MnTe was among the first materials with a clear experimental validation of such a broken Kramers degeneracy through angle-resolved photoemission spectroscopy measurements~\cite{osumi2024,lee2024,krempasky2024}. The remarkable splitting of magnons with opposite chirality, predicted to occur in altermagnets~\cite{Smejkal2023,Maier2023}, was also first observed experimentally in $\alpha$-MnTe~\cite{liu2024}.

Despite the large interest in MnTe, a comprehensive microscopic description of its low-energy magnetic properties is still lacking. Existing studies have estimated the dominant intersite exchange interactions (IEI) either by fitting experimentally measured inelastic neutron scattering (INS) spectra~\cite{szuszkiewicz2006,liu2024} or through energy-mapping approaches~\cite{mu2019,mazin2023,rooj2024,alaei2025} based on density functional theory (DFT) supplemented with effective on-site Coulomb repulsion, i.~e. DFT+U method. DFT-based approaches~\cite{alaei2025,sandratskii2025,wu2025,kravchuk2025} have also successfully reproduced the altermagnetic splitting of magnon branches at intermediate and large wave vectors. However, these IEI calculations generally neglect the exchange anisotropy arising from spin-orbit coupling (SOC) as well as other possible sources for the two-site anisotropy. Consequently, the microscopic origin of the experimentally observed easy-plane anisotropy~\cite{kriegner2017} has not been so far identified, with semi-empirical single-ion terms employed in the literature~\cite{alaei2025,kravchuk2025}. Moreover, although a recent DFT-based study has computed the magnon branch splitting near $\Gamma$ as a function of the single-ion anisotropy strength~\cite{kravchuk2025}, no fully first-principles calculation have so far reproduced the experimentally observed behavior of the magnon modes in this region. 

Beyond their role in the magnon spectrum, the IEI in $\alpha$-MnTe are strongly coupled to the lattice. Upon the onset of the long-range magnetic order the material exhibits a remarkably large magnetostrictive volume contraction as observed by X-ray diffraction~\cite{baral2023}. The coupling of its altermagnetic electronic structure to pressure has also been recently investigated both theoretically and experimentally~\cite{devaraj2024,carlisle2025}. However, the microscopic origin of the unusually large magnetostriction effect in $\alpha$-MnTe has not been yet addressed theoretically.

Here we address these open questions by evaluating a fully \textit{ab initio} effective low-energy Hamiltonian for $\alpha$-MnTe. To that end we employ the DFT + dynamical mean field theory (DMFT) framework~\cite{georges1996,anisimov1997,Lichtenstein1998,aichhorn2016} within the Hubbard-I approximation~\cite{hubbardI} to calculate the paramagnetic electronic structure of $\alpha$-MnTe. The force theorem in Hubbard-I (FT-HI)~\cite{pourovskii2016} is then applied as post-processing to evaluate the IEI. Solving the resulting effective Hamiltonian at the mean-field (MF) level, we find that, although several mechanisms contribute to the easy-plane anisotropy of $\alpha$-MnTe, it is dominated by the classical dipole-dipole interaction (DDI). Evaluating magnetic excitations in the ground-state ordered phase, we reproduce the observed magnitude of the altermagnetic splitting at large wave vectors~\cite{liu2024}. Furthermore, we identify classical DDI as the dominant origin of the magnon gap at $\Gamma$, providing \textit{ab initio} evidence that the mechanism recently proposed for a model altermagnet~\cite{jin2025} is realized in $\alpha$-MnTe. Finally, we evaluate the effect of pressure on the IEI and the magnetic excitation spectra. The calculated magnetostrictive volume compression and its temperature evolution are in quantitative agreement with experiment and point out to a strong magneto-volume coupling; however, we find no significant impact of pressure on the magnon altermagnetic gap.

\paragraph*{Electronic structure calculations.} In our charge self-consistent DFT+DMFT calculations, the DFT part was carried out using the linear augmented plane-wave (LAPW) WIEN2k package~\cite{wien2k} employing the local density approximation~\cite{lda} for the exchange-correlation functional, a dense 12$\times$12$\times$6 grid for the Brillouin zone sampling and the LAPW basis cutoff parameter $R_{MT}K_{max}=8.0$, where $R_{MT}$ is the smallest muffin-tin radius and $K_{max}$ is the largest wave vector in use.

The DMFT part was performed using the TRIQS library~\cite{triqs,aichhorn2016} with the charge self-consistency implemented in accordance with Ref.~\cite{aichhorn2011}. The Mn $3d$ Wannier orbitals were constructed using the projective approach of Ref.~\cite{aichhorn2009} from the Kohn-Sham bands enclosed by the energy window [$-1.5$; $+2.7$] eV relative to the DFT Fermi energy. The DMFT impurity problem on the Mn $3d$ shell was solved within the Hubbard-I approximation. The on-site Coulomb repulsion energy $U$ and the Hund's exchange coupling $J_H$ were set to 6.0 and 1.0 eV respectively, in reasonable agreement with the findings of previous computational studies on $\alpha$-MnTe~\cite{mu2019,youn2004}. A more detailed discussion on this parameter choice is provided in Sec.~I of the supplementary material~\cite{supplmat}.

\paragraph*{Low-energy effective Hamiltonian.} The converged DFT+DMFT electronic structure was subsequently used as input for the FT-HI method, implemented in the MagInt software~\cite{magint}. The FT-HI method evaluates full intersite kinetic exchange, including anisotropic and higher-rank multipolar contributions, in strongly correlated insulators by applying infinitesimal two-sites symmetry-breaking charge fluctuations within the relevant atomic ground-state multiplet, which is, in the present case, the Mn $3d^5$ Hund's rule sextet $^6S_{5/2}$. Combining the calculated IEI with the single-ion terms obtained in converged DFT+DMFT calculations and the classical DDI between local magnetic moments we obtained an effective low-energy Hamiltonian $H_{\mathrm{eff}}$ of the following form

\begin{equation}
H_{\mathrm{eff}} = \sum_{\langle ij\rangle}\sum_{\alpha\beta} \tilde{J}_{\alpha\beta}(\vR_{ij}) S^{\alpha}_i S^{\beta}_j  + H_{\mathrm{si}}
\label{eq:1}
\end{equation}
where $\langle ij\rangle$ runs over all Mn-Mn bonds, $S^{\alpha}_i$ and $S^{\beta}_j$ are the spin-5/2 operators of the projection $\alpha,\beta=x$, $y$, or $z$ acting on the Mn sites $\mathbf{R}_i$ and $\mathbf{R}_j$, respectively, $\tilde{J}_{\alpha\beta}(\vR_{ij})$ is the intersite coupling for the bond $\vR_{ij}=\vR_j-\vR_i$ . The latter consists of an exchange contribution supplemented by a dipolar one

\begin{equation}
\tilde{J}_{\alpha\beta}(\vR_{ij})  =J_{\alpha\beta}(\vR_{ij}) + J^{DDI}_{\alpha\beta}(\vR_{ij}),
\label{eq:2}
\end{equation}
where $J_{\alpha\beta}(\vR_{ij})$ is the IEI and $J^{DDI}_{\alpha\beta}(\vR_{ij})$ is the classical magnetic DDI between spin-5/2 magnetic moments evaluated in the standard way, see, e.~g., Ref.~\cite{blundell2001}. Finally, $H_{\mathrm{si}}$ is the single-ion term acting at each site $\mathbf{R}_i$, which arises due to SOC admixing excited multiplets into the pure spin-5/2 state. It is given by the splitting of the $^6S_{5/2}$ ground-state multiplet in the converged DFT+DMFT calculations within Hubbard-I approximation. This splitting comes out to be quite small in $\alpha$-MnTe, with its total width equal to 60 $\mu eV$ as shown in Fig.~\ref{fig:iei}c.

In principle, SOC also induces IEI between all high rank multipoles that can be hosted by the ground-state Mn $^6S_{5/2}$ sextet. We have also evaluated those multipolar IEI in $\alpha$-MnTe and found them to be negligible (being all below 1 $\mu$eV, see Suppl.~Sec.~II~\cite{supplmat}) and not leading to any discernible effects in the magnetic order or excitations. Correspondingly, we neglect multipolar IEI in Eq.~\eqref{eq:1}.

In contrast, anisotropic IEI contributions are physically significant. To quantify them we decompose the exchange matrices $J(\vR_{ij})$ into isotropic and anisotropic terms

\begin{equation}
J_{\alpha\beta}(\vR_{ij})=J_n\delta_{\alpha\beta}+\epsilon_{\alpha\beta\gamma}D_{\gamma}(\vR_{ij})+\Gamma_{\alpha\beta}(\vR_{ij}),
\label{eq:3}
\end{equation}
where the terms on the right hand side are the isotropic IEI $J_n=\frac{1}{3}\text{Tr}[J(\vR_{ij})]$ labeled by its nearest-neighbor (NN) shell index $n$, $\vR_{ij}\in n$, the antisymmetric exchange anisotropy defined by the Dzyaloshinskii-Moriya (DM) vector $\mathbf{D}(\vR_{ij})$ with components $D_{\gamma}(\vR_{ij})=\frac{1}{2}\sum_{\alpha\beta}\epsilon_{\gamma\alpha\beta}J_{\alpha\beta}(\vR_{ij})$ and the symmetric exchange anisotropy in terms of $\Gamma_{\alpha\beta}(\vR_{ij})=\frac{1}{2}[J_{\alpha\beta}(\vR_{ij})+J_{\beta\alpha}(\vR_{ij})]-J_n$. The magnitude of the IEI anisotropy for a given NN shell is quantified by the length of the DM vector $|\mathbf{D}(\vR_{ij})|$ as well as by $|\Gamma|=\sqrt{\mathrm{Tr}[\Gamma^2(\vR_{ij})|}$.

The calculated isotropic and anisotropic IEI are displayed in Figs.~\ref{fig:iei}d,~\ref{fig:iei}e and~\ref{fig:iei}f, respectively. In particular, the isotropic contributions $J_n$ are found to be dominated by the NN AFM IEI $J_1$. The calculated $J_n$ parameters for the first several coordination shells are in a good agreement with recent experimental estimates~\cite{liu2024} and DFT+U total energy approaches~\cite{alaei2025,mu2019,mazin2023,rooj2024,krause2013}. We also obtain a non-negligible altermagnetic splitting of the isotropic IEI $J_{10}$ and $J_{10}'$ for the two inequivalent $10^{\text{th}}$ NN exchange paths displayed in Figure~\ref{fig:iei}a. Their difference $\Delta J_{10}$  amounts to 0.13 meV and thus it agrees reasonably well with the value 0.09 meV extracted from fitting experimental magnon dispersions~\cite{liu2024}, while being noticeably larger than the earlier first-principles DFT+U estimate~\cite{alaei2025} of 0.04 meV. 

Our calculations, however, predict an AFM $2^{\text{nd}}$ NN coupling $J_2$, in contrast to the value extracted from those INS measurements~\cite{liu2024}. We note that only first three $J_n$ couplings together with the tenth NN ones were employed in the fit. In contrast, we find AFM IEI up to $\sim 0.1$ meV for distant bonds in the coordination shells from $4^{\text{th}}$ to $9^{\text{th}}$. Although small individually, these couplings contribute substantially in total, owing to the large number of bonds in these shells.

The anisotropic contributions to the exchange couplings are significantly smaller than their isotropic counterparts. As shown in Fig.~\ref{fig:iei}f, the antisymmetric DM interaction for the $3^{\text{rd}}$ NN shell is clearly the leading anisotropic term, while the symmetric anisotropic IEI $\Gamma$ is of a comparable magnitude to the classical DDI prefactor $\lambda_{DDI}=\frac{g^2\mu_0\mu_B^2}{4\pi|R^{(n)}|^3}$, where $g=2$ is the Land\'e  factor for the Mn $^6S_{5/2}$ multiplet, $\mu_0$ is the vacuum permeability and $R^{(n)}$ is the length of $n^{\text{th}}$ NN Mn-Mn connecting vectors.

\paragraph*{Magnetic order and magnetic anisotropy.} The MF solution of the resulting low-energy effective Hamiltonian, obtained using the McPhase package together with an in-house module~\cite{mcphase}, correctly reproduces the experimentally observed planar transverse AFM order shown in Fig.~\ref{fig:iei}a below $T_N\simeq520$~K, with magnetic moments of $\sim5$~$\mu_B$ lying within the ($a,b$) plane. The agreement with the experimental magnetic structure~\cite{kunitomi1964,kriegner2017} supports the validity of our interaction parameters and their ability to describe the altermagnetic phase of $\alpha$-MnTe. We note that this magnetic order is not destabilized by an AFM $J_2$; a different magnetic configuration becomes stable only if $J_2\gtrsim2J_3$, since the $2^{\text{nd}}$ NN coordination number is half that of the $3^{\text{rd}}$ NN shell. The MF overestimation of experimental $T_N$ by about 70\% is within the range expected for 3D systems.

To identify the origin of the in-plane magnetic anisotropy of $\alpha$-MnTe~\cite{kriegner2017} reproduced by our \textit{ab initio} effective Hamiltonian, we performed a series of MF calculations in which all anisotropic terms in~\eqref{eq:1} were switched off selectively: the single-ion anisotropy, the classical DDI and the anisotropic intersite exchange. Our data show that each of them alone is sufficient to reproduce the experimentally observed moment orientation, therefore no single interaction can be considered decisive.

To establish which mechanism dominates, we instead fixed the N\'eel vector orientation while evaluating the corresponding total magnetic energy contributions. For an arbitrary value for the azimuthal angle $\theta$ within the $xz$ plane (see Fig.~\ref{fig:iei}b), we define the corresponding magnetic moment operator

\begin{equation}
\hat{m}(\theta) = -g(\hat{S}_x \cos\theta + \hat{S}_z \sin\theta)
\label{eq:4}
\end{equation}
where $\hat{S}_a$ are the spin-5/2 operators for the projection $a=x, z$. Diagonalizing  $\hat{m}(\theta)$ one finds the eigenvector $\psi(\theta)$ corresponding to the largest eigenvalue, i.~e., to the fully saturated moment along the $\theta$ direction. The most negative eigenvalue corresponds to the oppositely saturated moment, i.~e., $\psi(\theta+\pi)$.

The constrained N\'eel-vector state $\Psi(\theta)$ of the magnetic unit cell was subsequently constructed as a product of the single-site wavefunctions $\psi(\theta)$ and $\psi(\theta+\pi)$, corresponding to the two opposite moment directions. The different contributions to the total magnetic energy for each $\theta$ were then calculated as expectation values with respect to this state, i.~e.

\begin{equation}
E_p(\theta)
=
\langle\Psi(\theta)|
\hat{H}_{p}
|\Psi(\theta)\rangle,
\label{eq:5}
\end{equation}
where $\hat{H}_p$ denotes the individual terms of $H_{\mathrm{eff}}$ in Eq.~\eqref{eq:1}.

\begin{figure}[t]
    \centering
    \includegraphics[width=\linewidth]{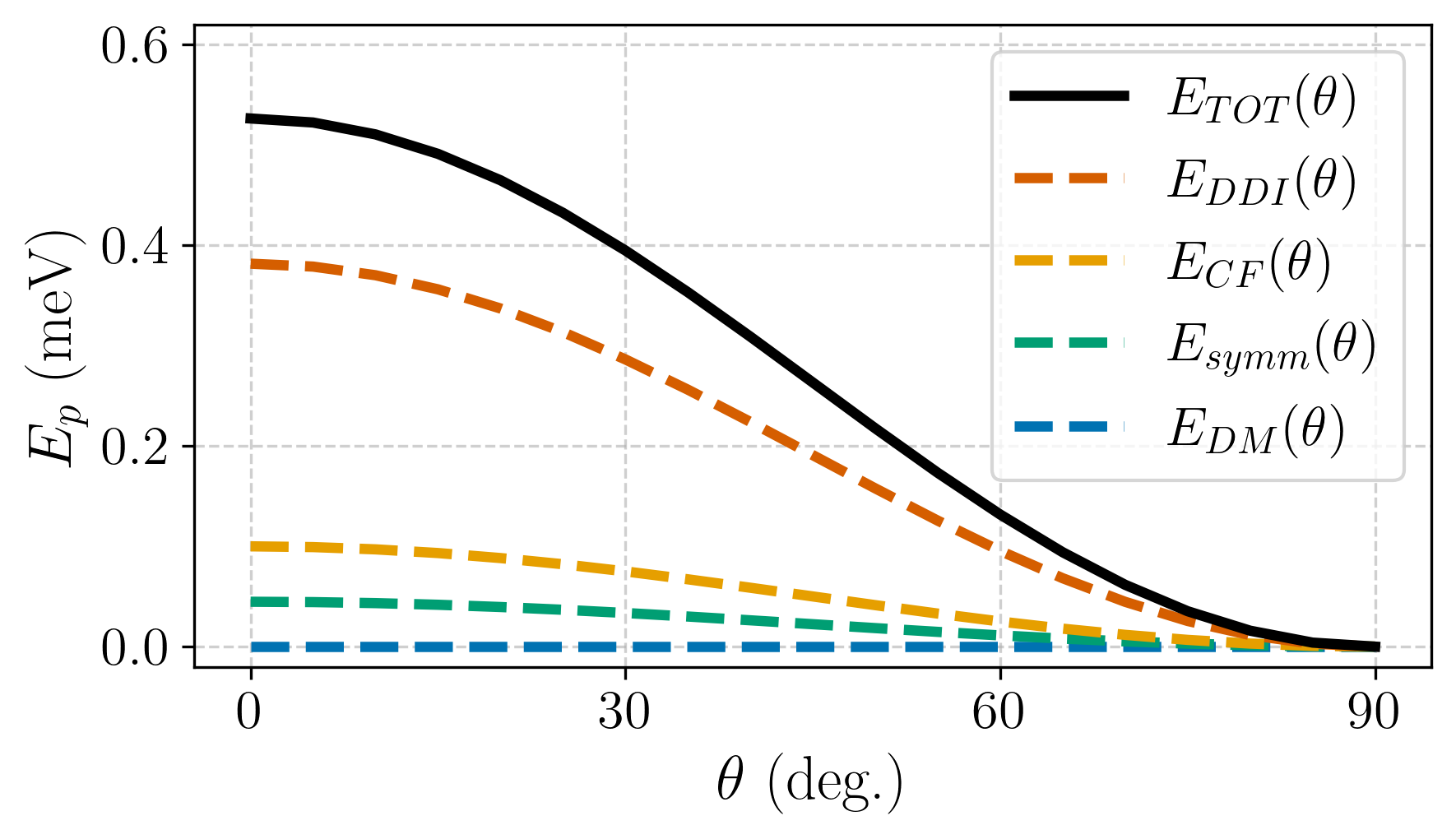}
    \caption{Total energy contributions in the effective Hamiltonian of Eq.~\eqref{eq:1} as a function of the azimuthal angle $\theta$ of the N\'eel vector: Dzyaloshinskii-Moriya (DM), symmetric exchange (symm), single-ion crystal field (CF) and classical dipole-dipole interaction (DDI). Each contribution is plotted relative to its value for the ground-state planar orientation ($\theta=90^{\circ}$). }
    \label{fig:aniso}
\end{figure}

The resulting angular energy profiles are plotted in Fig.~\ref{fig:aniso} and reveal a clear hierarchy among the different contributions to the in-plane magnetic anisotropy. The classical DDI provides the largest contribution, followed by the crystal-field (CF) interaction and the symmetric anisotropic exchange, while, surprisingly, the DM interaction has a negligible impact.

A similar procedure can be employed to investigate the dependence of the magnetic anisotropy on the polar angle. However, due to the very small energy scales involved in this case, i.~e. around $10^{-2}$ $\mu$eV, the targeted energy profiles could not be reliably resolved within the numerical accuracy of our approach.

\paragraph*{Magnetic excitations.} In order to evaluate magnetic excitations in $\alpha$-MnTe from the \textit{ab initio} effective Hamiltonian of Eq.~\eqref{eq:1}, we first calculated its dynamical susceptibility $\chi(\vq,E)$ in the MF ground-state within the standard random phase approximation (see, e.~g.~\cite{Jensen1991}):

\begin{equation}
\bar{\chi}^{-1}(\vq,E)=\bar{\chi}^{-1}_0(E)+\bar{J}_{\vq},
\label{eq:6}
\end{equation}
where $\chi_0(E)$ is the single-site susceptibility evaluated from the on-site MF  potential in the ordered phase, $J_{\vq}$ is the Fourier-transformed intersite couplings $\tilde{J}(\vR_{ij})$. The overbar $\bar{ }$ denotes a matrix in the combined index of the magnetic-unit cell site and spin projection $\alpha$. The dynamical structure factor $S(\mathbf{q},E)$ measured in INS is then obtained from the absorptive part of the generalized susceptibility $\bar{\chi}''(\mathbf{q},E)=-\frac{i}{2}[\bar{\chi}(\mathbf{q},E)-\bar{\chi}^{\dagger}(\mathbf{q},E)]$ by the fluctuation-dissipation theorem~\cite{Jensen1991}.

Finally, the measured INS intensity is computed from $S(\mathbf{q},E)$ as follows

\begin{equation}
I(\vq,E) \propto F^2(|\vq|)\sum_{\alpha\beta}(\delta_{\alpha\beta}-q_\alpha q_\beta)  S_{\alpha\beta}(\mathbf{q},E), 
\label{eq:7}
\end{equation}
where $\mathbf{q}$ ($E$) is the momentum (energy) transfer to the system, $F(|\mathbf{q}|)$ is the form-factor~\footnote{In our INS simulations, we employed angular-dependent form-factors for the Mn $^6S_{5/2}$ including all multipolar contributions following the formalism of Ref.~\cite{pourovskii2021}. However, since the corrections beyond the dipole approximation are tiny in the case of $\alpha$-MnTe, for simplicity we write the form-factors in Eq.~\eqref{eq:7} in this approximation.}.

\begin{figure}[b]
    \centering

    \begin{subfigure}{\linewidth}
        \centering
        \begin{overpic}[width=\linewidth]{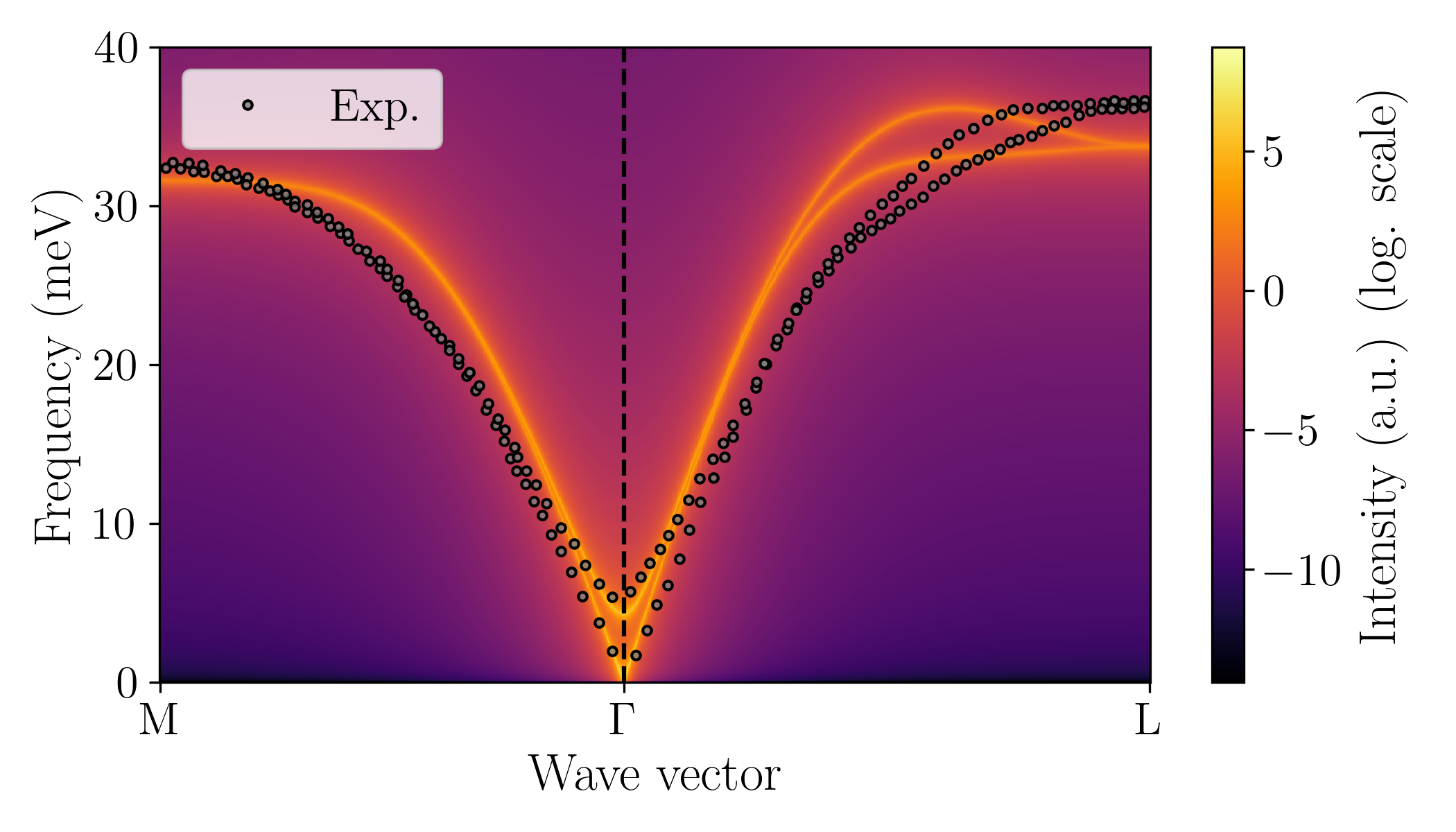}
        \put(0,55){(a)}
        \end{overpic}
    \end{subfigure}

    \begin{subfigure}{0.5\linewidth}
        \centering
        \raisebox{1.0cm}{
        \begin{overpic}[width=\linewidth]{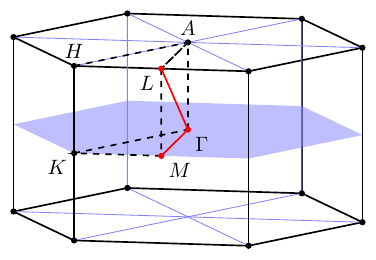}
        \put(0,70){(b)}
        \end{overpic}
        }
    \end{subfigure}
    \hfill
    \begin{subfigure}{0.45\linewidth}
        \centering
        \begin{overpic}[width=\linewidth]{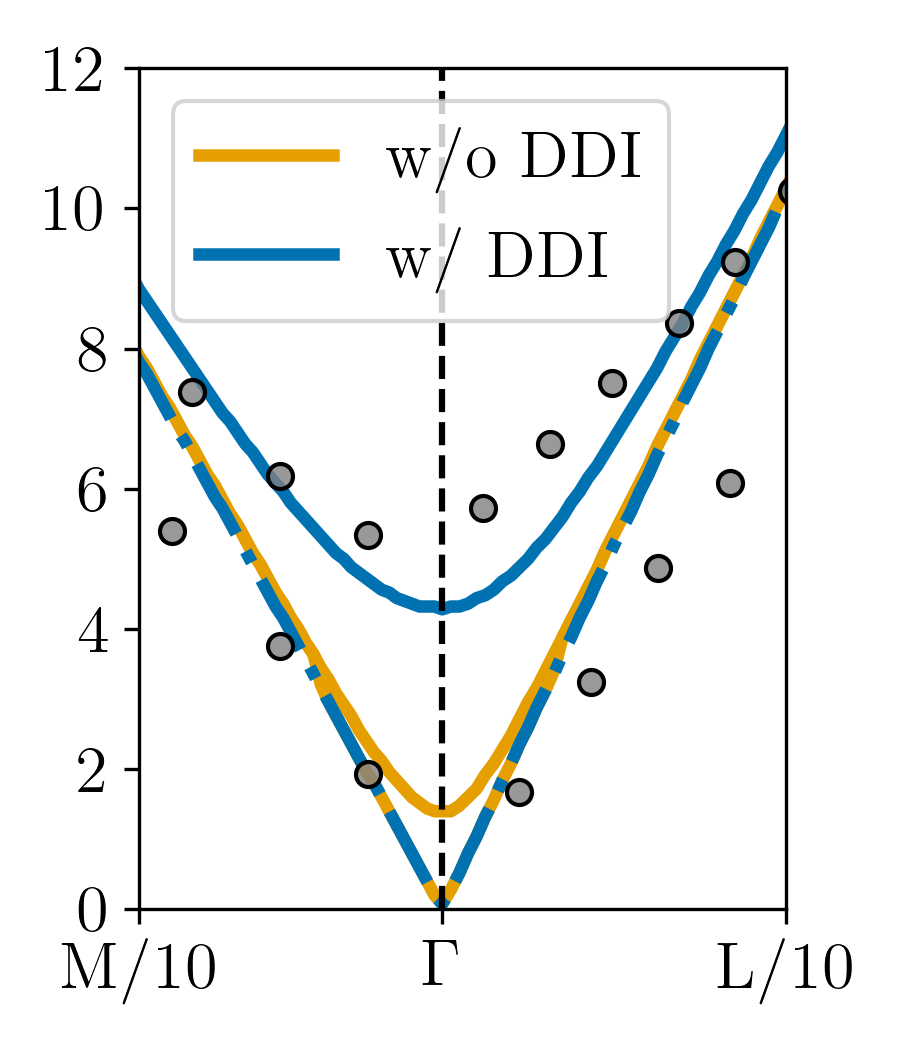}
        \put(-5,90){(c)}
        \end{overpic}
    \end{subfigure}

    \caption{(a) Calculated inelastic neutron-scattering spectra of $\alpha$-MnTe along the M-$\Gamma$-L path in the first Brillouin zone; black dots denote the experimental data of Ref.~\cite{liu2024}. (b) First Brillouin zone of $\alpha$-MnTe showing the high-symmetry points, the altermagnetic nodal planes (light blue) and the selected path (red). (c) Close-up of the magnon modes near $\Gamma$, blue and orange lines show the magnon dispersions calculated with and without the classical dipole–dipole interactions (DDI); black dots are experimental data~\cite{liu2024}.}
    \label{fig:magnexc}
\end{figure}

The plot in Fig.~\ref{fig:magnexc}a shows the calculated INS intensity along the M–$\Gamma$–L path (see Fig.~\ref{fig:magnexc}b) in the first Brillouin zone exhibiting sharp magnon dispersions. The first segment in the selected path (i.~e. M–$\Gamma$) lies within the altermagnetic nodal plane perpendicular to the reciprocal lattice vector $k_c$ along the M–$\Gamma$ segment, while the second part (i.~e. $\Gamma$–L) does not intersect any altermagnetic nodal plane.

The calculated spectrum clearly indicates that the two magnon branches exhibit the same anisotropic features of the altermagnetic spin splitting as found in the electronic bandstructure of $\alpha$-MnTe~\cite{lee2024}. Namely, their degeneracy at large wave vectors is protected only in the altermagnetic nodal planes. A clear altermagnetic splitting is observed along the antinodal path $\Gamma$-L and its magnitude is in a good agreement with experimental INS data~\cite{liu2024}.

We find that, while the weak classical DDI plays a minor role at larger wave vectors, it significantly modifies the spectrum near the $\Gamma$ point (see Fig.~\ref{fig:magnexc}c). There, it shifts the upper magnon branch by about 3 meV and enhances the splitting of the two magnon branches by a factor of about 3.2, closely matching the experimentally observed dispersion~\cite{liu2024}. A non-zero splitting of the two magnon modes is present even in the absence of classical DDI due to a combined effect of symmetric anisotropic exchange and CF interaction.

\paragraph*{Magnetostriction and pressure effects.} To evaluate the magnetostrictive volume compression in the ordered phase we first calculated the evolution of the effective Hamiltonian of Eq.~\eqref{eq:1} vs pressure. The crystal structures corresponding to  different pressures were extracted from the neutron powder diffraction data reported in Ref.~\cite{carlisle2025} for T = 350 K and P = 0.0, 0.3, 0.5, 1.0 GPa.  For a detailed discussion on the calculated IEI pressure dependence see Suppl.~Sec.~III~\cite{supplmat}. 

The resulting effective Hamiltonians were solved using the same MF approach as above. The calculated magnetic MF free energy, which varies linearly with volume, was supplemented with the elastic term calculated from the Birch–Murnaghan equation of state~\cite{birch1947,murnaghan1944,poirier2000} for the paramagnetic phase, parametrized by the bulk modulus $B_0=47.3$ GPa and its pressure derivative $B_0'=3.0$ from Ref.~\cite{paszkowicz1997} and the temperature-dependent equilibrium volume $V_0(T)$ predicted by Debye-Grüneisen model function~\cite{debye1912,gruneisen1912,chatterji2011}.
The resulting total free energy was minimized to identify the new equilibrium volume $V^*(T)$ in the presence of magnetic order, with the corresponding magnetostrictive volume compression evaluated as $\frac{\Delta V(T)}{V_0(T)} = \frac{V^*(T)-V_0(T)}{V_0(T)}$.

\begin{figure}[h]
\centering
\includegraphics[width=\linewidth]{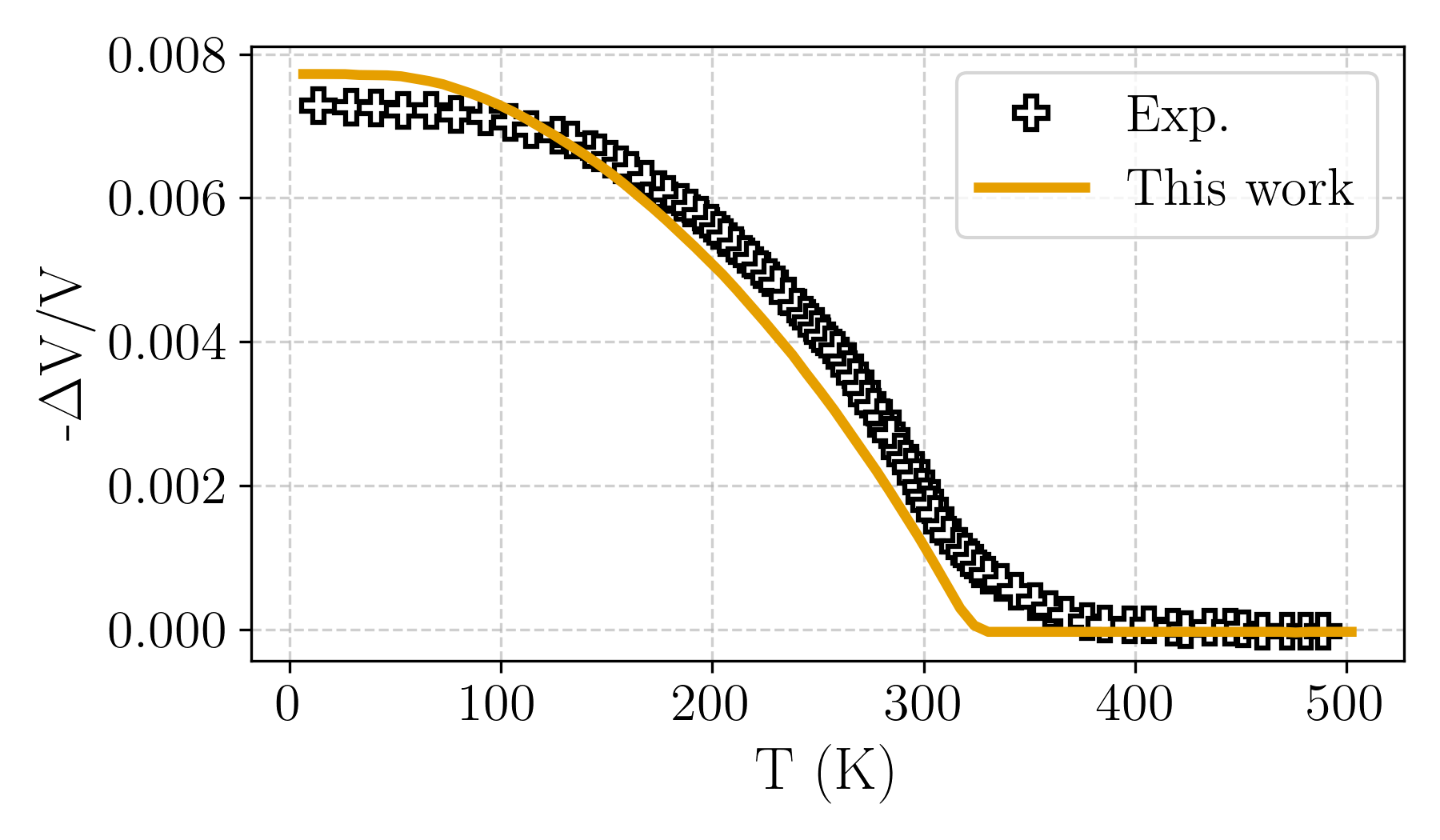}
\caption{Temperature dependence of the computed magnetostriction shift, compared with X-ray diffraction data~\cite{baral2023}. The temperature axis of the theoretical curve has been rescaled to account for the $T_N$ overestimation due to the mean-field approximation.}
\label{fig:magnstr}
\end{figure}

The estimated magnetostriction compression is shown in Figure~\ref{fig:magnstr} together with recent experimental data~\cite{baral2023} over the whole temperature range below 500 K. The calculations predict a very sizeable magnetostriction that is consistent with the reference data both in terms of magnitude at low-T limit and temperature dependence. The magneto-volume coupling is thus quantitatively well captured within the present theoretical framework that is essentially based on kinetic IEI captured by our \textit{ab initio} approach.

The magnetostriction in $\alpha$-MnTe is actually much larger than the typical range in other Mott insulators, e.~g., transition-metal fluorides like -7.72$\times10^{-4}$ for CrF${}_2$, -1.16$\times10^{-4}$ for CuF${}_2$, -1$\times10^{-3}$ for MnF${}_2$ and -4.5$\times10^{-4}$ for NiF${}_2$, pointing out to a high sensitivity of Mn-Te hybridization, and, correspondingly, of superexchange, to pressure.

The latter may suggest that the magnetic excitation spectrum could also be strongly sensitive to pressure. We have carried out extensive calculations to explore the effect of uniform pressure on the magnetic anisotropy and excitations in $\alpha$-MnTe. As detailed in the Suppl.~Sec.~III~\cite{supplmat}, while the pressure sizeably affect both the altermagnetic magnon splitting and the magnetic dispersions at $\Gamma$, it does not lead to any qualitative modifications with respect to the zero-pressure excitation spectrum. The easy-plane anisotropy is robust upon the changes of $c/a$ within the explored pressure range up to 15 GPa.

\paragraph*{Conclusions.} 
Our study provides a robust qualitative and quantitative description of the low-energy physics of $\alpha$-MnTe, based on the \textit{ab initio} determination of an effective spin Hamiltonian incorporating exchange interactions, local CF terms and classical DDI. The estimated exchange parameters confirm that the altermagnetic phase of $\alpha$-MnTe is governed primarily by the isotropic contributions to the IEI for the first three NN shells. Furthermore, they consistently stabilize the experimentally observed magnetic ground state within the MF approximation. Building upon this effective model, we investigate the microscopic origin of the easy-plane magnetic anisotropy of $\alpha$-MnTe. We show that while all interactions favor in-plane magnetic anisotropy, their contributions exhibit a clear hierarchy, with DDI dominating, followed by CF and symmetric anisotropic exchange, while the DM interaction has a negligible effect. The magnetic excitation spectrum calculated from the {\it ab initio} effective Hamiltonian captures both the altermagnetic splitting of the magnon branches at large wave vectors and the magnon gap and branch splitting at $\Gamma$. The features at $\Gamma$ are mainly induced by classical DDI. We further find that the unusually large magnetostriction compression in $\alpha$-MnTe is fully accounted for by the pressure dependence of the intersite exchange. Although pressure significantly affects the energetics of the altermagnetic phase, it does not induce qualitative changes in either the magnetic anisotropy or the structure of the excitation spectrum.

\paragraph*{Acknowledgments.} This work benefited from government funding managed by the French National Research Agency under the France 2030 program, reference number ‘ANR-22-EXES-0013’. D.~F.~M. thanks the Austrian Science Fund (FWF) grant DOI: 10.55776/PAT2121825. The authors thank the computational facilities of the Austrian Scientific Computing (ASC). 

\bibliographystyle{apsrev4-2-pagesdoi}
\bibliography{apssamp}

\end{document}


\preprint{APS/123-QED}

\title{Supplementary material for 'Spin Hamiltonian, magnetic anisotropy and \\ excitation spectrum in altermagnetic $\alpha$-MnTe from first principles'}

\date{\today}

\maketitle

\onecolumngrid

\begin{center}

\vspace{-0.5cm}

\large Matteo Costa

\textit{CPHT, CNRS, École polytechnique, Institut Polytechnique de Paris, 91120 Palaiseau, France}
\\ \textit{and Collège de France, Université PSL, 11 place Marcelin Berthelot, 75005 Paris, France}

\vspace{0.6cm}

Dario Fiore Mosca

\textit{University of Vienna, Faculty of Physics, Center for Computational\\ Materials Science, Kolingasse 14-16, 1090, Vienna, Austria}

\vspace{0.6cm}

Cesare Franchini

\textit{University of Vienna, Faculty of Physics, Center for Computational\\ Materials Science, Kolingasse 14-16, 1090, Vienna, Austria}
\\ \textit{and Department of Physics and Astronomy \enquote{Augusto Righi},\\ Alma Mater Studiorum - Università di Bologna, Bologna, 40127 Italy}

\vspace{0.6cm}

Leonid V. Pourovskii

\textit{CPHT, CNRS, École polytechnique, Institut Polytechnique de Paris, 91120 Palaiseau, France}
\\ \textit{and Collège de France, Université PSL, 11 place Marcelin Berthelot, 75005 Paris, France}

\end{center}

\section{Relevance of the on-site Coulomb repulsion energy}
\label{sec:sm_coulomb}

Our choice for the on-site Coulomb repulsion energy $U$ and the Hund's exchange coupling $J_H$ is consistent with previous computational studies, where estimates for these parameters were obtained using both linear response theory~\cite{mu2019} and a constrained-DFT-type method~\cite{youn2004} based on specific Slater's transition states rules. Nevertheless, we note that smaller values of the effective coupling ($U_{eff}=U-J_H=$ 3 to 4 eV) were also used in the literature~\cite{kriegner2017,krause2013,alaei2025,mazin2023,rooj2024,sandratskii2025,carlisle2025,devaraj2024}. For this reason, we examine the sensitivity of our results to variations of $U$ in this section.

Employing the same DFT+DMFT combined approach to the paramagnetic electronic structure for several values of $U$ in the range [4; 8] eV, and then applying the FT-HI method as detailed in the main text, we investigate the dependence of the exchange interactions of $\alpha$-MnTe on $U$. As expected from superexchange theory, increasing $U$ leads to a noticeable reduction in the magnitude of the IEI. This trend is illustrated in Suppl. Fig.~\ref{fig:sm_ieis_vs_u} and it can be observed for both the isotropic and the anisotropic IEI.

\begin{figure}[h]
    \centering

    \begin{subfigure}{0.355\linewidth}
        \centering
        \includegraphics[width=\linewidth]{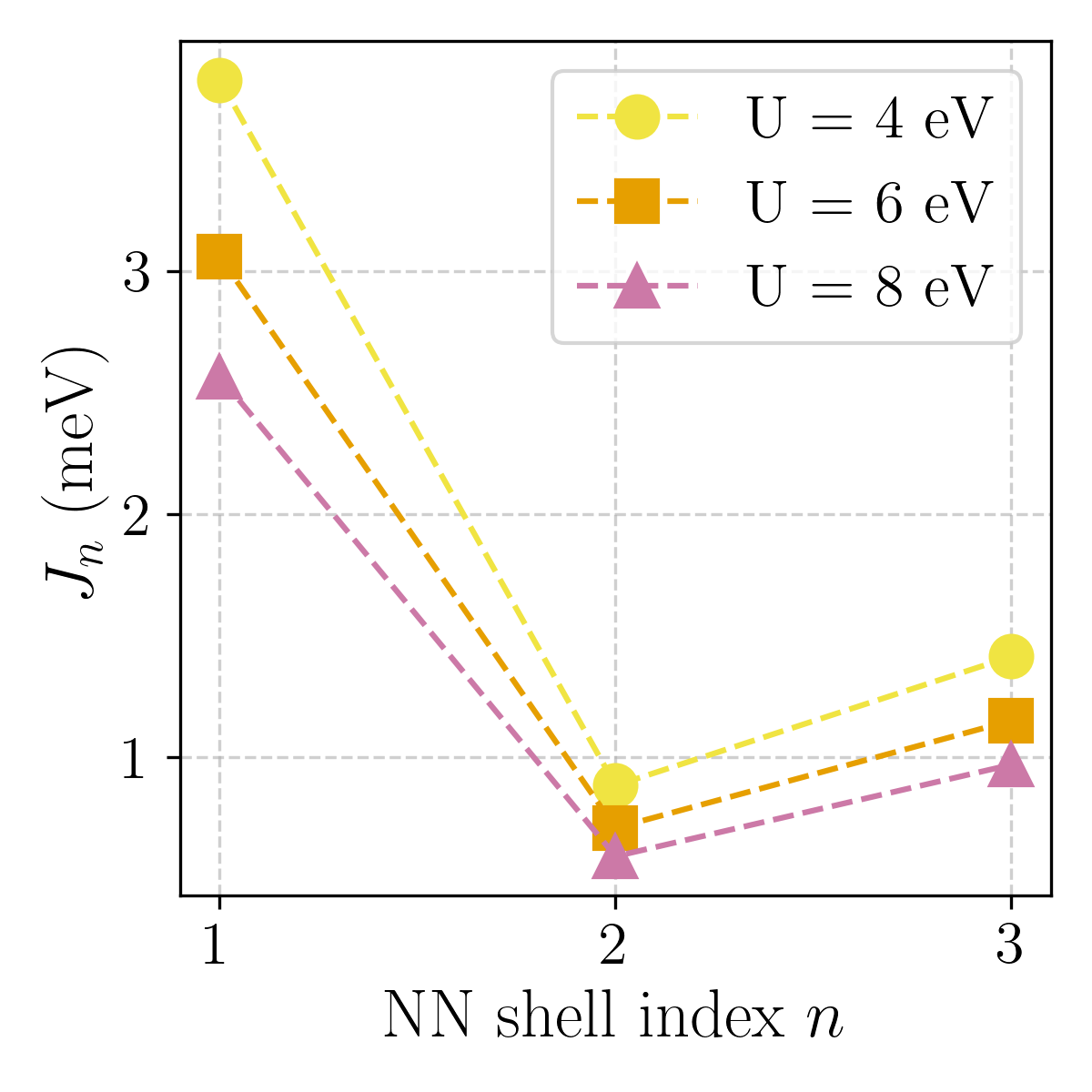}
        \put(-180,170){\large(a)}
    \end{subfigure}
    \begin{subfigure}{0.53\linewidth}
        \centering
        \includegraphics[width=\linewidth]{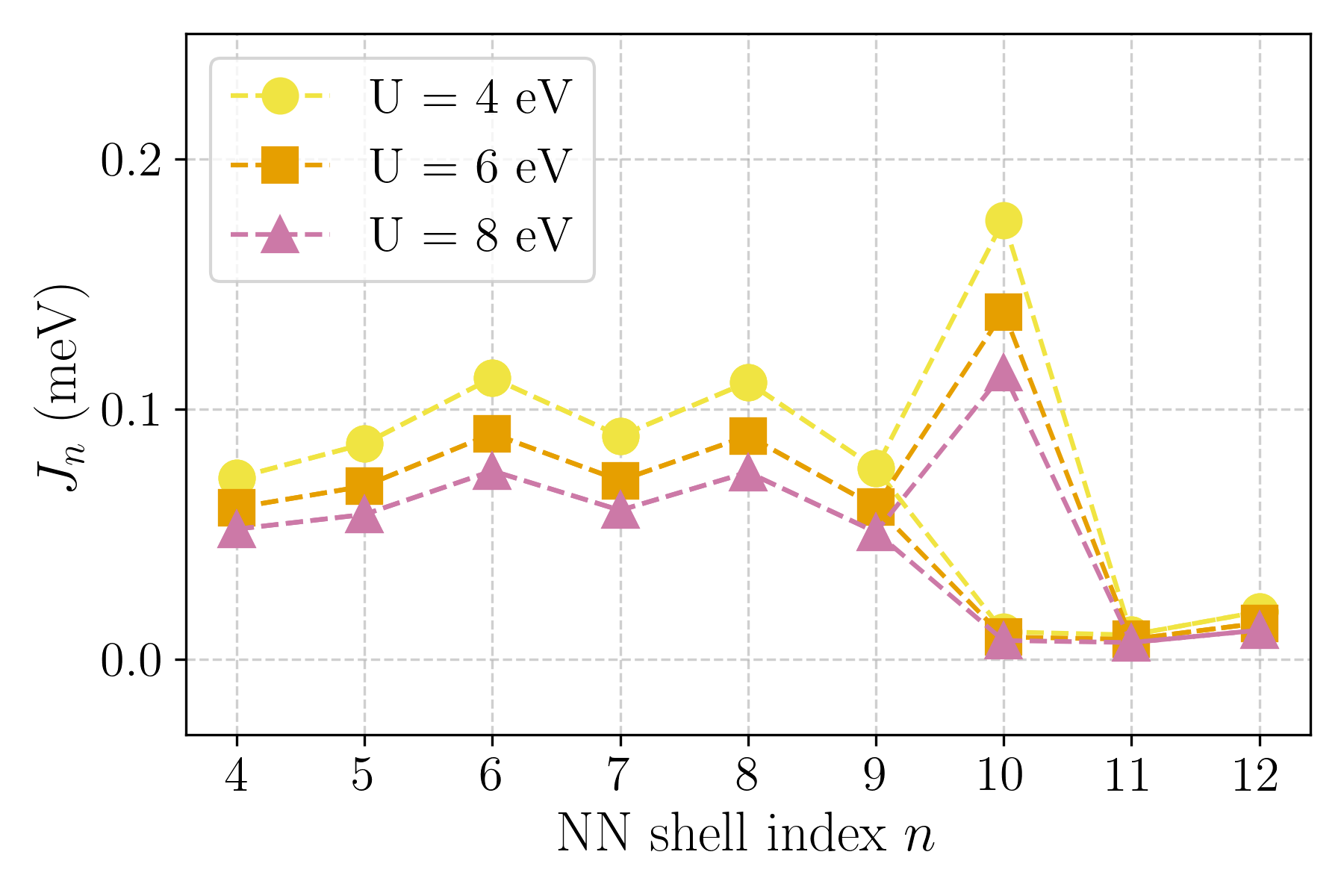}
        \put(-270,170){\large(b)}
    \end{subfigure}
\end{figure}
\begin{figure}[h]
    \ContinuedFloat
    \centering
        
    \begin{subfigure}{0.9\linewidth}
        \centering
        \includegraphics[width=\linewidth]{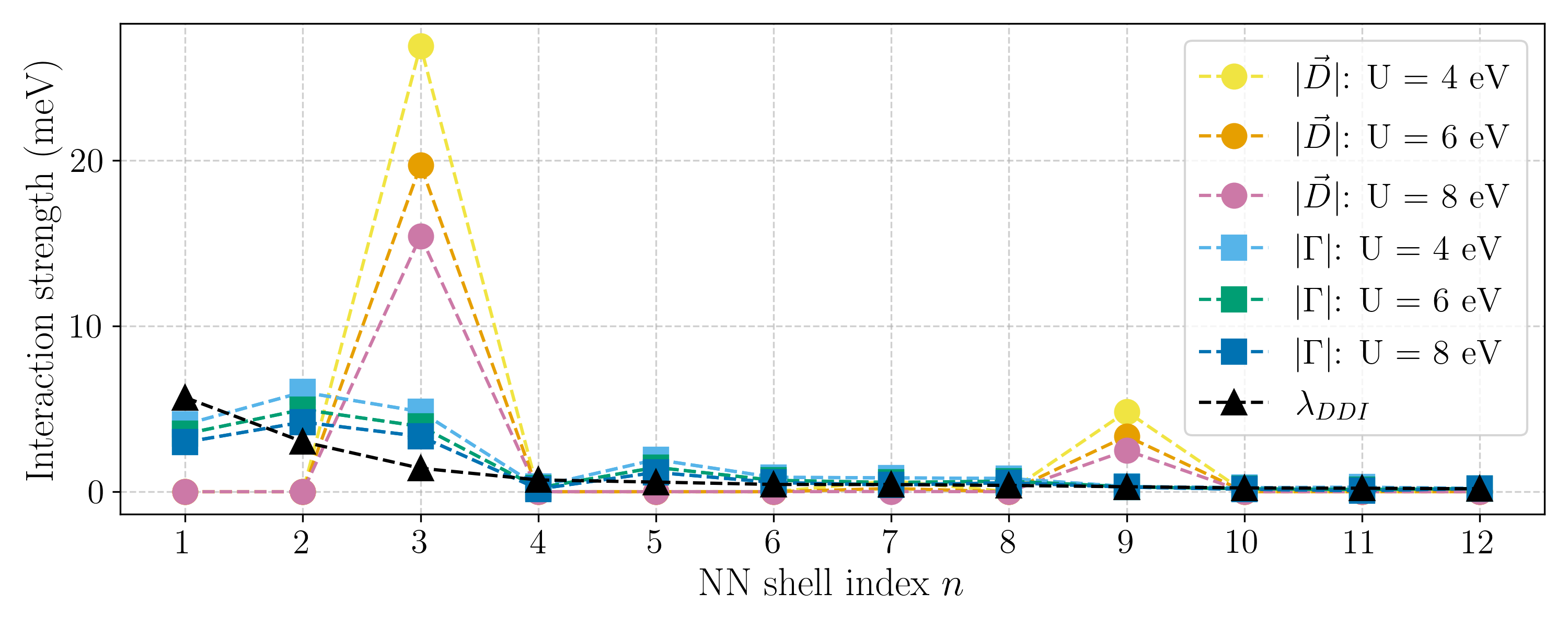}
        \put(-462,180){\large(c)}
    \end{subfigure}

    \caption{Isotropic contributions to the IEI for the first three (a) and $4^{\text{th}}$-to-$12^{\text{th}}$ (b) NN shells at U = 4, 6, 8 eV. (c) Magnitude of the symmetric anisotropic exchange and the DM interaction for the first twelve NN shells at U = 4, 6, 8 eV, together with the radial prefactor $\lambda_{DDI}$ of the classical DDI.}
    \label{fig:sm_ieis_vs_u}
\end{figure}

However, the qualitative behavior of the exchange interactions remain essentially unchanged across the considered range of $U$. In particular, the overall hierarchy of the IEI is preserved.

\begin{figure}[h]
    \centering
    \includegraphics[width=0.7\linewidth]{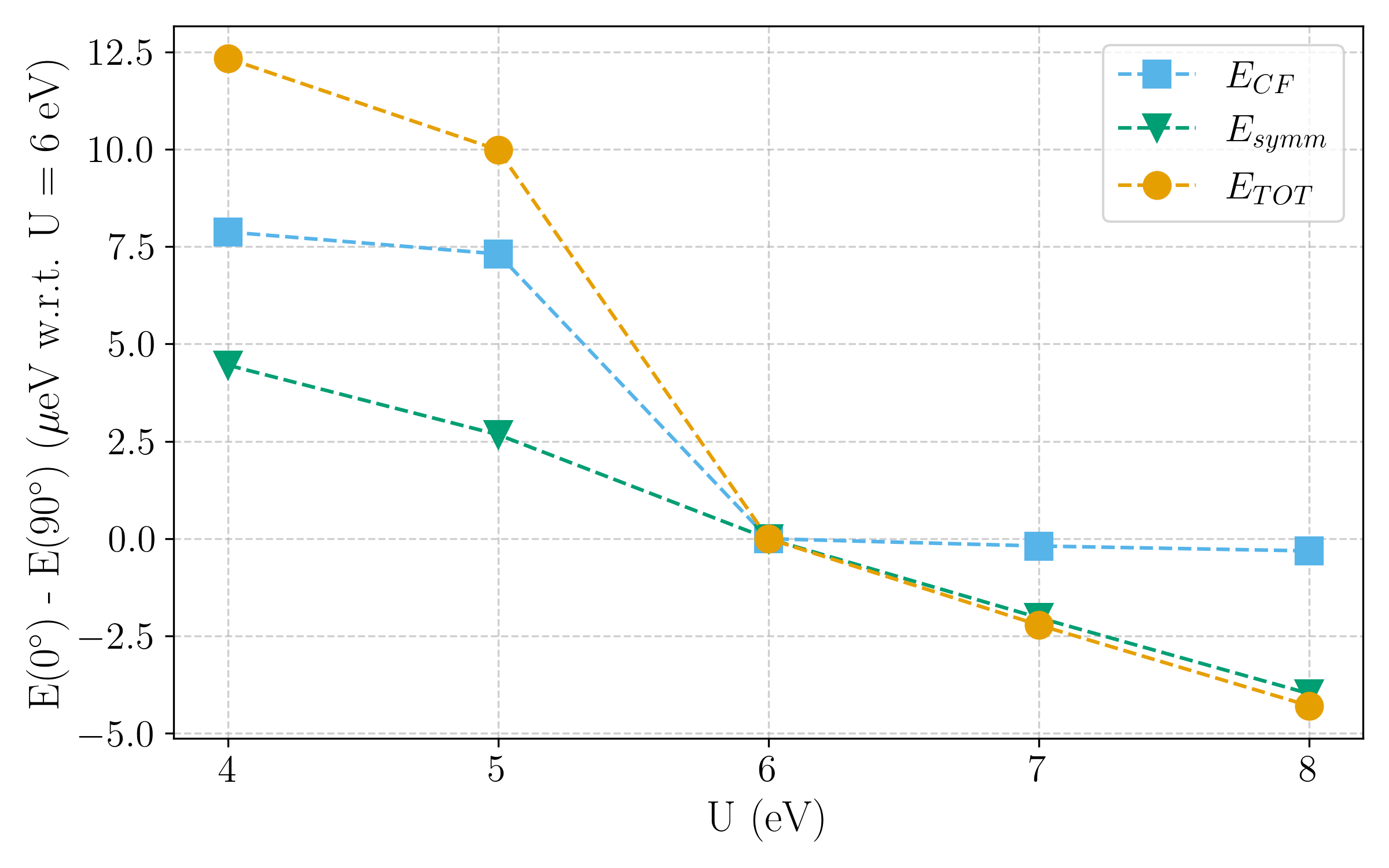}
    \caption{Dependence of the total magnetic anisotropy energy (orange) and its contributions from the CF interaction (light blue) and symmetric anisotropic exchange (green) on $U$, shown relative to their respective values at $U$ = 6 eV.}
    \label{fig:sm_mae_vs_u}
\end{figure}

For each value of $U$, we solve the resulting effective Hamiltonian in MF and apply the same approach described in the main text to investigate the dependence of the magnetic anisotropy energy on $U$. Since the classical DDI are not affected by the choice of $U$ and the DM interactions keep on not contributing to the magnetic anisotropy (see Fig. 2 in the main text), we focus our attention on the energy contributions arising from CF and symmetric anisotropic exchange. The resulting trends are depicted in Suppl. Fig.~\ref{fig:sm_mae_vs_u}, showing that they both exhibit a decreasing behavior with increasing $U$.

\vspace{1.5cm}

In particular, the decrease of the symmetric anisotropic exchange contribution consistently reflects the corresponding reduction of the symmetric part of the anisotropic IEI, as shown in Suppl. Fig.~\ref{fig:sm_ieis_vs_u}c. Nevertheless, the magnitude of these variations remains very small, of the order of $\mu$eV, when compared with the total magnetic anisotropy energy of approximately 0.53 meV. We can therefore conclude that the planar magnetic anisotropy of $\alpha$-MnTe is only weakly affected by the choice of $U$ within the physically reasonable range considered here.

\begin{figure}[h]
    \centering
    \includegraphics[width=0.7\linewidth]{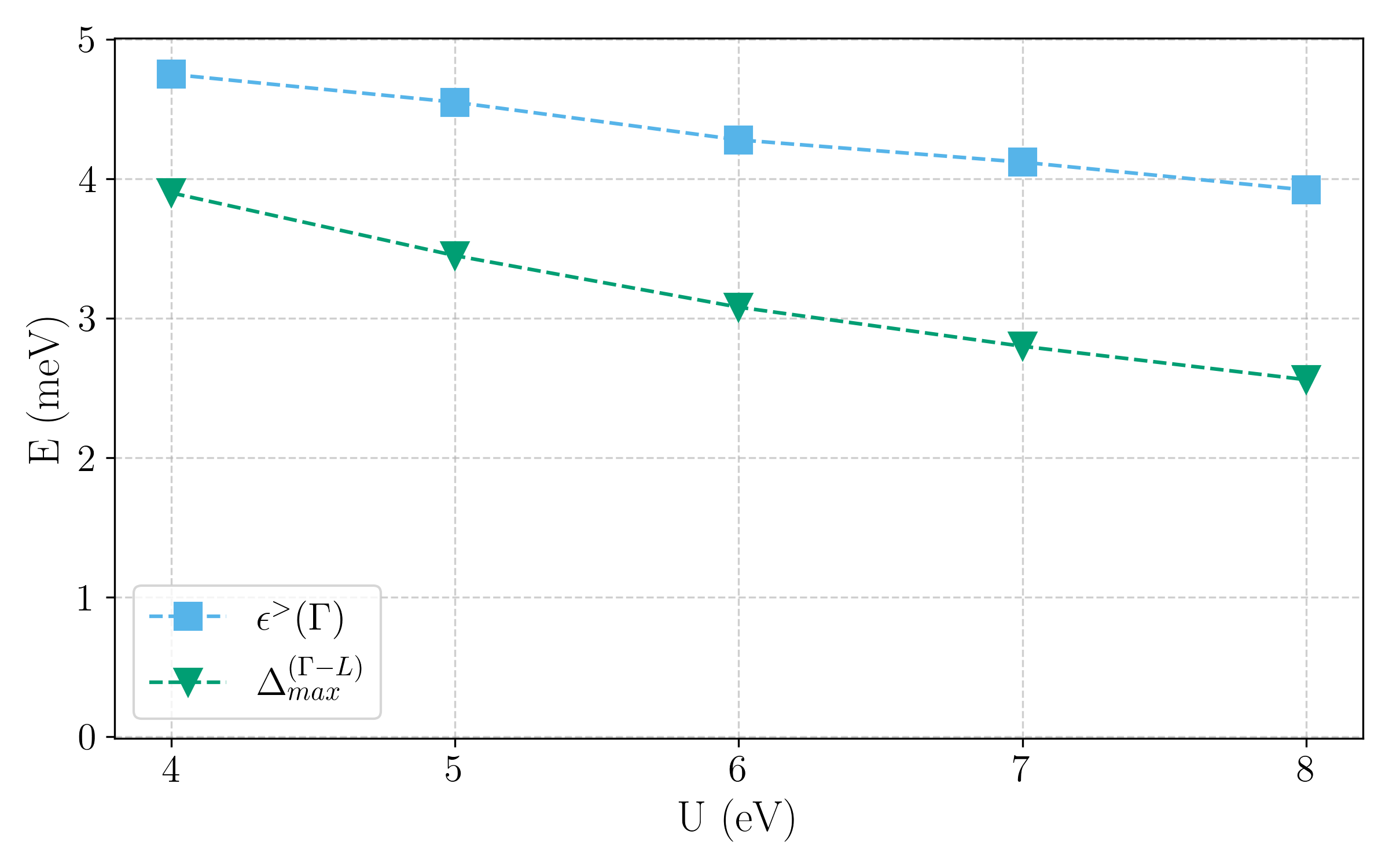}
    \caption{Upper magnon-branch energy $\epsilon^{>}$ at the $\Gamma$ point (light blue) and maximum magnon splitting $\Delta^{(\Gamma-L)}_{max}$ along the $\Gamma$-L high-symmetry path (green) at several values for $U$.}
    \label{fig:sm_magnsplits_vs_u}
\end{figure}

Finally, we calculate the magnetic excitation spectrum along the M-$\Gamma$-L path at each value of $U$, following the approach detailed in the main text. As shown in Suppl. Fig.~\ref{fig:sm_magnsplits_vs_u}, both the energy of the upper magnon branch $\epsilon^{>}$ at $\Gamma$ and the maximum magnon splitting at large wave vectors along the $\Gamma$-L segment decrease with increasing $U$. While the first trend can be attributed to the reduction of the symmetric part of the anisotropic IEI with increasing $U$ (see Suppl. Fig.~\ref{fig:sm_ieis_vs_u}c), the latter originates from the decreasing difference $\Delta J_{10}$, which is clearly visible in Suppl. Fig.~\ref{fig:sm_ieis_vs_u}b. Although the characteristic magnon features of $\alpha$-MnTe exhibit a non-negligible $U$ dependence, we report no physically significant qualitative change in its magnetic excitation spectra.

\section{Multipolar contributions to the IEIs}
\label{sec:sm_multipolar}

To assess the importance of higher-order multipolar contributions to the IEI in $\alpha$-MnTe, we consider the full exchange Hamiltonian

\begin{equation}
\hat{H}_{IEI} = \sum_{\langle ij\rangle}\sum_{KK'QQ'} V_{KK'}^{QQ'}(\vR_{ij}) \hat{O}_{K}^{Q}(\vR_i) \hat{O}_{K'}^{Q'}(\vR_j).
\end{equation}
expressed in terms of the Hermitian spherical tensors~\cite{Santini2009} $\hat{O}_K^Q(\vR_i)$ for $J=5/2$ with rank $K=1,2,3$ and projection $Q=-K,-K+1,\dots,+K$, acting on Mn site at position $\vR_i$. The operators $\hat{O}_K^Q(\vR_i)$ are orthonormal, i.~e. $\mathrm{Tr}\left[\hat{O}_K^Q\cdot \hat{O}_{K'}^{Q'}\right]=\delta_{KK'}\delta_{QQ'}$. However, since all IEI contributions are reported in spin-operator notation in the main text, the spherical tensors must be consistently renormalized to $N=\sqrt{\text{Tr}[\hat{S}^z\cdot\hat{S}^z]}=\sqrt{35/2}$. So, defining the renormalized tensors as $\tilde{O}_K^Q(\vR_i)=N\hat{O}_K^Q(\vR_i)$, the full exchange Hamiltonian can be cast as

\begin{equation}
\label{eq:HIEI_full}
\hat{H}_{IEI} = \sum_{\langle ij\rangle}\sum_{KK'QQ'} V_{KK'}^{QQ'}(\vR_{ij}) \frac{\tilde{O}_{K}^{Q}(\vR_i)}{N} \frac{\tilde{O}_{K'}^{Q'}(\vR_j)}{N} = \sum_{\langle ij\rangle}\sum_{KK'QQ'} \tilde{V}_{KK'}^{QQ'}(\vR_{ij}) \tilde{O}_{K}^{Q}(\vR_i) \tilde{O}_{K'}^{Q'}(\vR_j)
\end{equation}
with the renormalized interaction parameter $\tilde{V}_{KK'}^{QQ'}(\vR_{ij})=V_{KK'}^{QQ'}(\vR_{ij})/N^2$. Furthermore, since $\tilde{O}_{1}^Q(\vR_i)$ coincide with the spin operators' components $\hat{S}_i^\alpha$ and an analogous correspondence holds between $\tilde{V}_{11}^{QQ'}(\vR_{ij})$ and $J_{\alpha\beta}(\vR_{ij})$, the Hamiltonian~\eqref{eq:HIEI_full}) can be thus decomposed into the leading DDI term and the weaker multipolar contribution as follows

\begin{equation}
\hat{H}_{IEI} = \sum_{\langle ij\rangle}\sum_{\alpha\beta} J_{\alpha\beta}(\vR_{ij}) S^{\alpha}_i S^{\beta}_j + \sum_{\langle ij\rangle}\sum_{\substack{KK'>1 \\ QQ'}} \tilde{V}_{KK'}^{QQ'}(\vR_{ij}) \tilde{O}_{K}^{Q}(\vR_i) \tilde{O}_{K'}^{Q'}(\vR_j).
\end{equation}

\begin{figure}[h]
    \centering
    \includegraphics[width=0.7\linewidth]{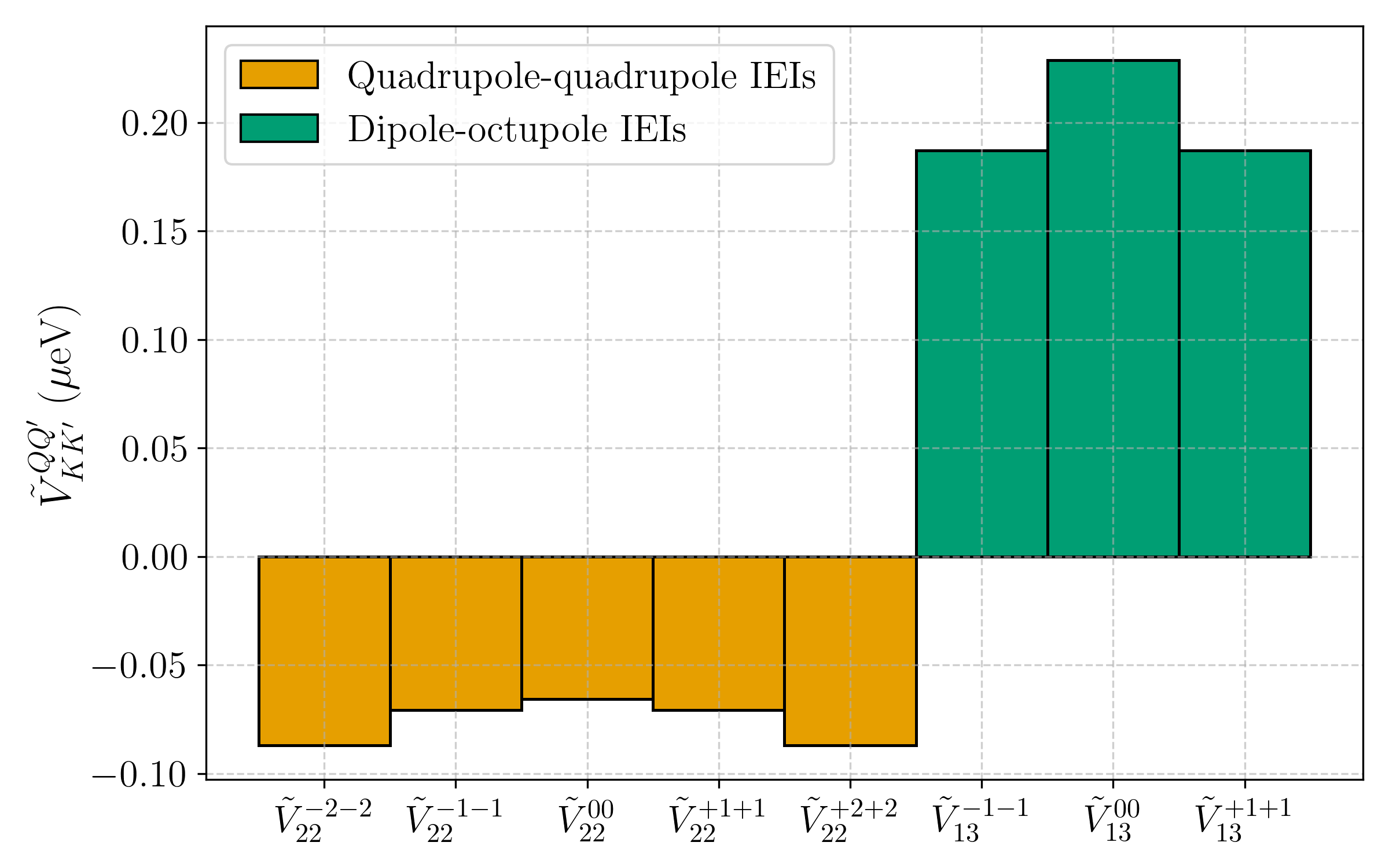}
    \caption{Histogram of the strongest multipolar contributions to the IEI in $\alpha$-MnTe, with proper renormalization to ensure their direct comparability with the exchange interactions reported in the main text. The largest values are associated with diagonal quadrupole–quadrupole and three specific dipole–octupole interactions for the $1^{\text{st}}$ NN bonds; however, none of them exceeds 1 $\mu$eV.}
    \label{fig:sm_multipolar}
\end{figure}

Among the resulting multipolar interactions, the largest values are found for three specific dipole–octupole interactions, i.~e. $\tilde{V}_{13}^{QQ}$ with $Q=-1,0,+1$, and the diagonal quadrupole–quadrupole terms, i.~e. $\tilde{V}_{22}^{QQ}$ with $Q=-2,-1,0,+1,+2$, for the Mn-Mn connecting vectors $\vR_{ij}$ in the $1^{\text{st}}$ NN shell. As shown in Suppl. Fig.~\ref{fig:sm_multipolar}, their magnitude in absolute value remains below 1 $\mu$eV. This indicates that higher-order multipolar interactions play only a minor role in the low-energy physics of $\alpha$-MnTe, when compared with the dominant exchange interactions discussed in the main text.

\section{Pressure effects}
\label{sec:sm_pressure}

To assess the effect of pressure on the magnetic properties of $\alpha$-MnTe, we perform the same \textit{ab initio} calculations described in the main text using the volume-dependent crystal structures from the neutron powder diffraction data reported in Ref.~\cite{carlisle2025} and the X-ray diffraction data in Ref.~\cite{wang2022}.

\begin{figure}[h]
    \centering

    \begin{subfigure}{0.355\linewidth}
        \centering
        \includegraphics[width=\linewidth]{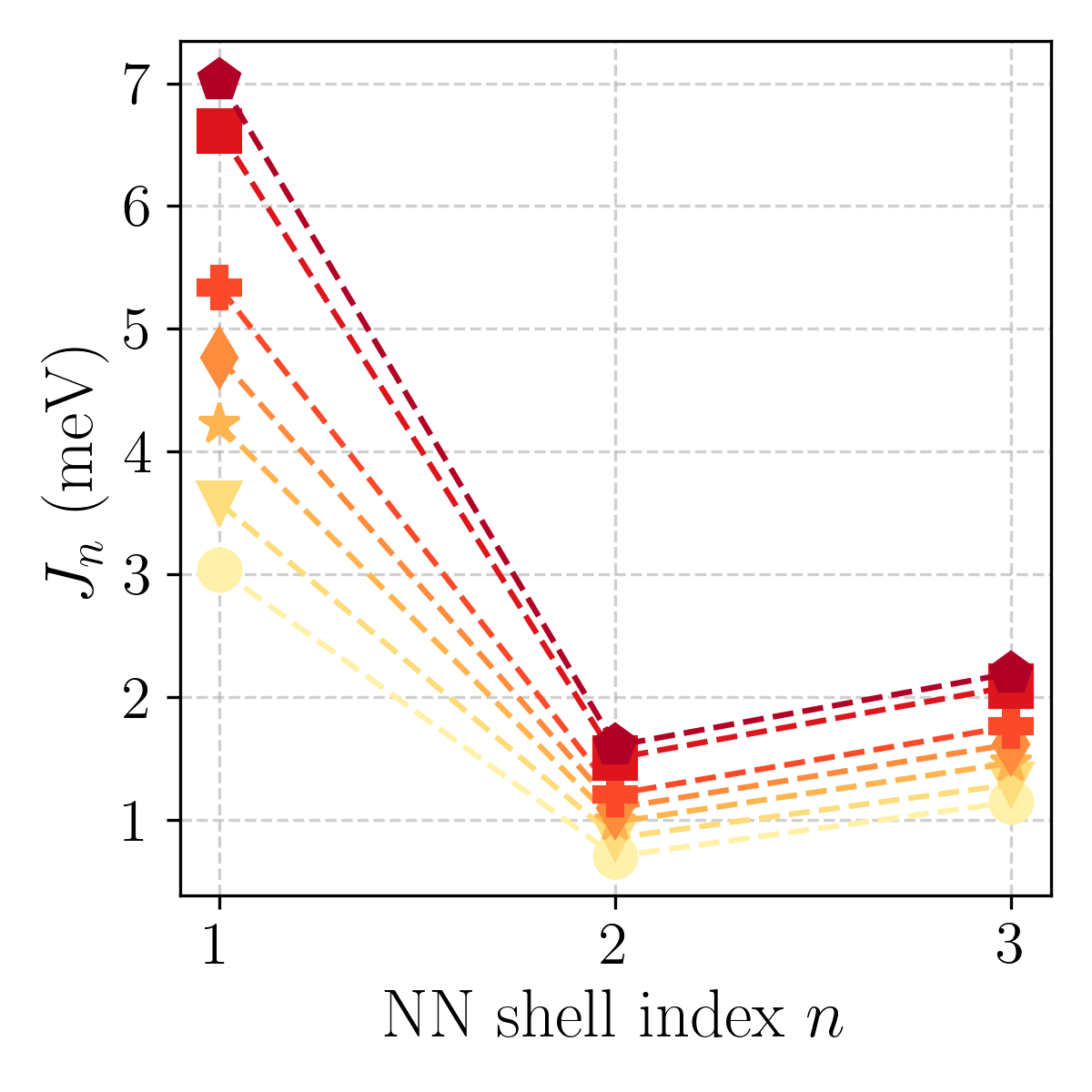}
        \put(-180,170){\large(a)}
    \end{subfigure}
    \begin{subfigure}{0.53\linewidth}
        \centering
        \includegraphics[width=\linewidth]{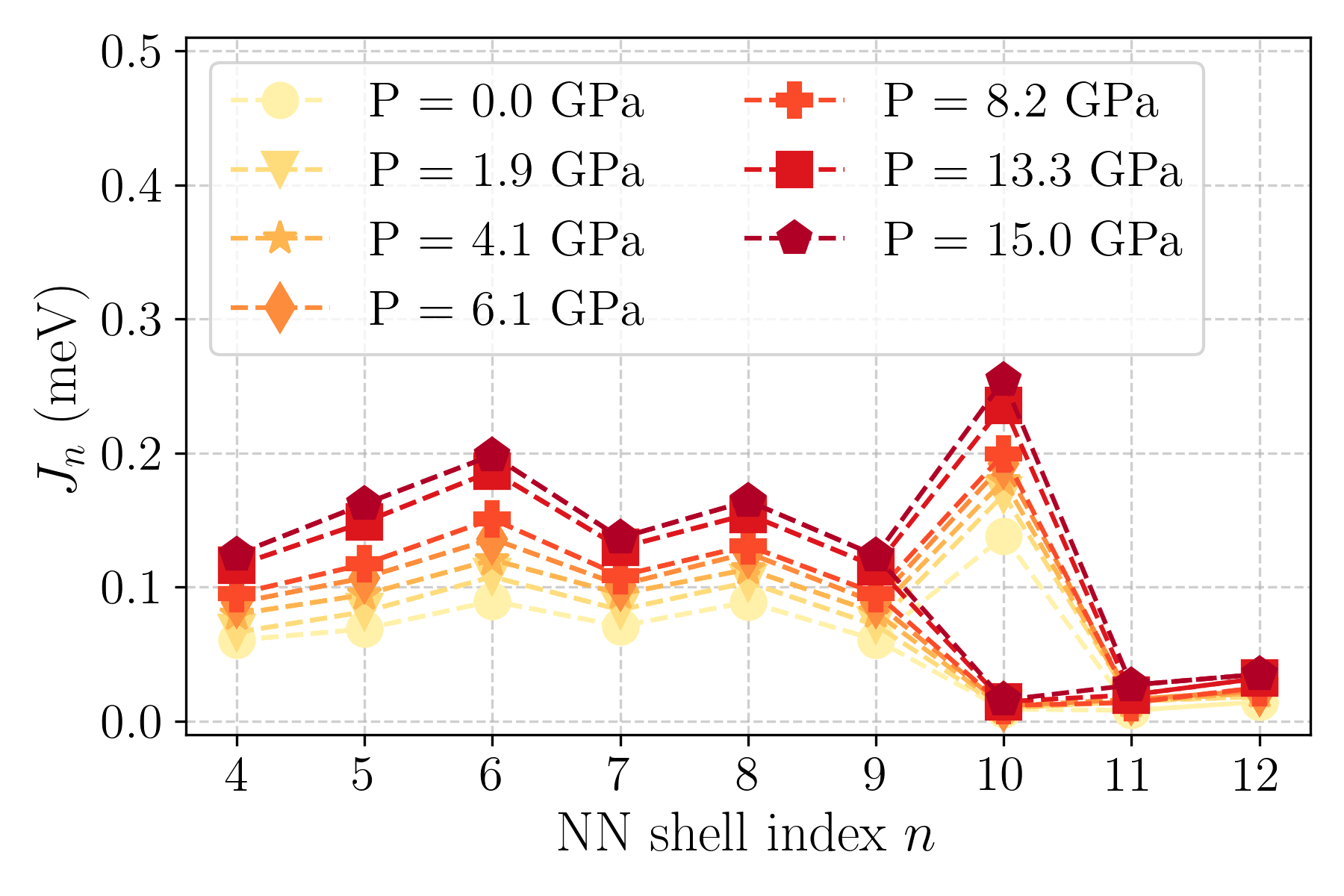}
        \put(-270,170){\large(b)}
    \end{subfigure}
    
    \begin{subfigure}{0.9\linewidth}
        \centering
        \includegraphics[width=\linewidth]{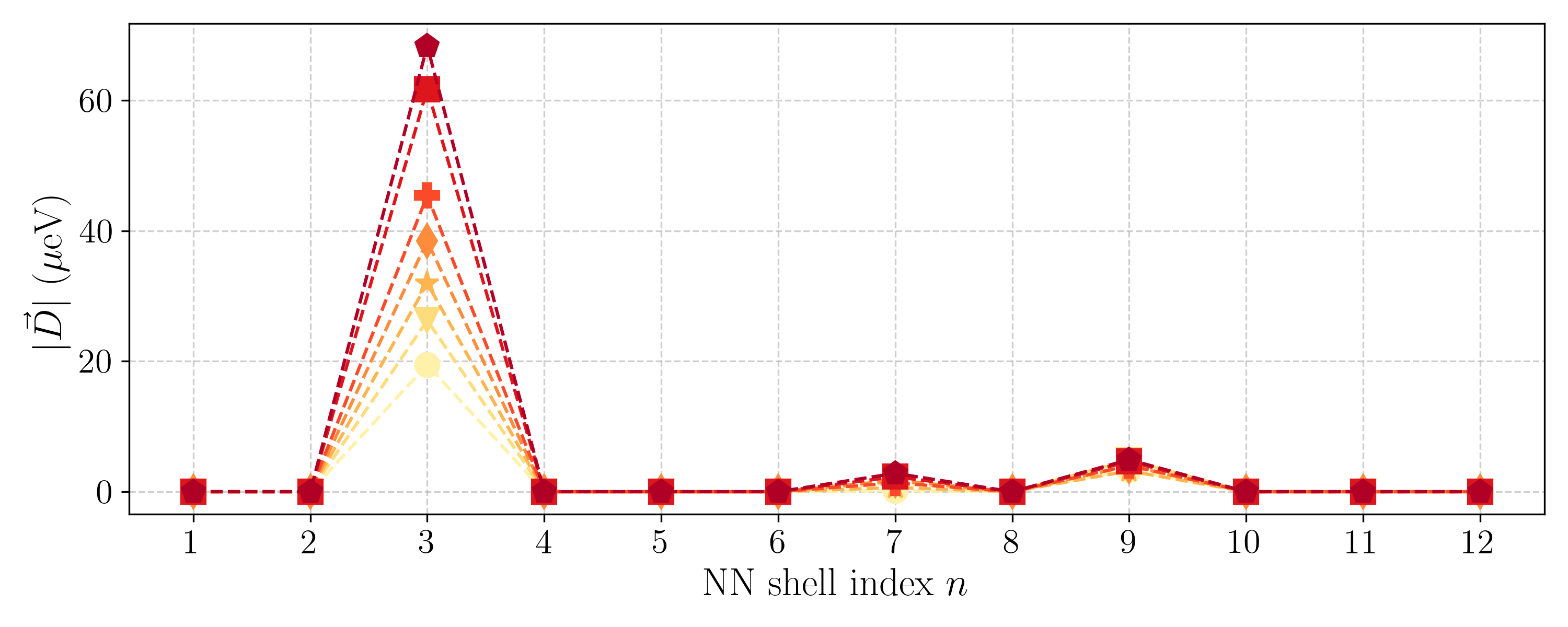}
        \put(-455,170){\large(c)}
    \end{subfigure}

    \begin{subfigure}{0.9\linewidth}
        \centering
        \includegraphics[width=\linewidth]{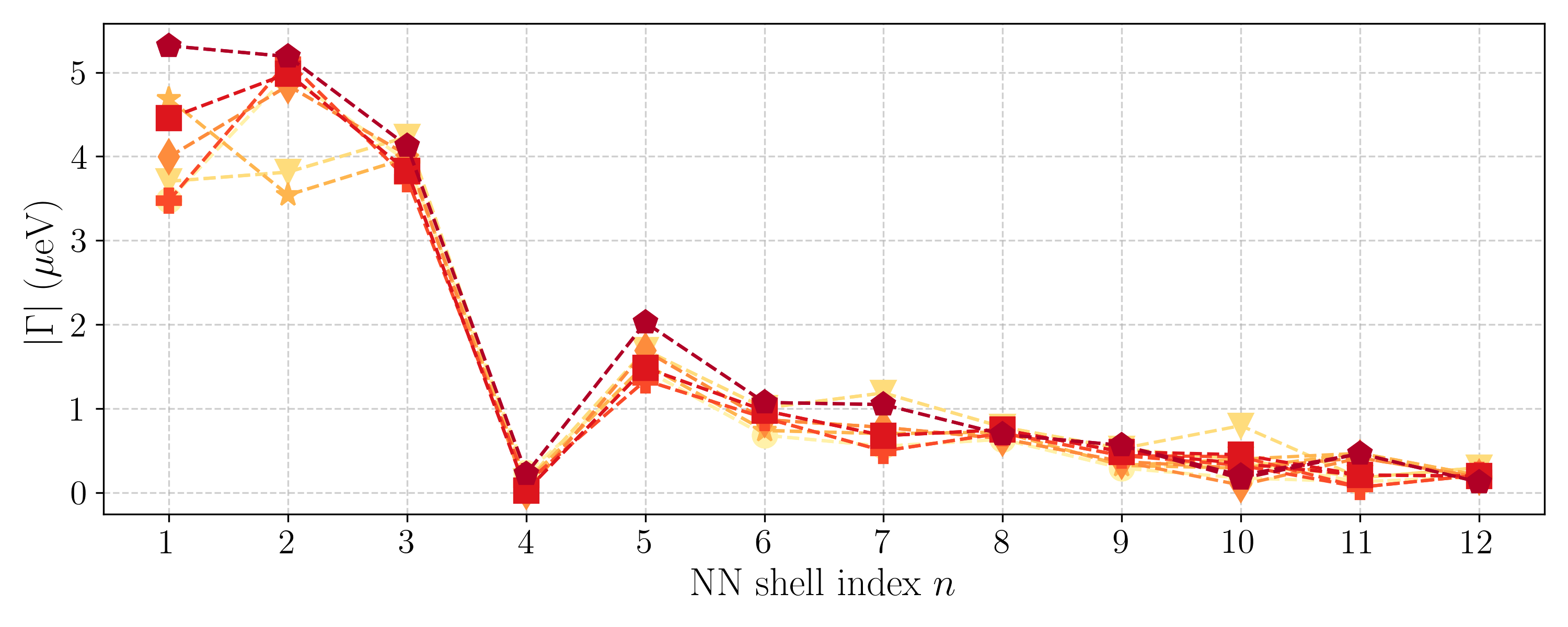}
        \put(-460,170){\large(d)}
    \end{subfigure}

    \caption{Effect of pressure on: the isotropic exchange interactions for the first three (a) and the $4^{\text{th}}$-to-$12^{\text{th}}$ (b) NN shells, the DM interaction (c) and the symmetric anisotropic exchange (d). The radial prefactor $\lambda_{DDI}$ is not shown since it reports negligible variations if compared with $|\vec{D}_3|$ or $|\Gamma_{1/2}|$.}
    \label{fig:sm_ieis_vs_p}
\end{figure}

The resulting pressure dependence of the isotropic and anisotropic IEI is shown in Suppl. Fig.~\ref{fig:sm_ieis_vs_p}. Overall, the isotropic contributions increase with pressure, consistently with the expected enhancement of superexchange interaction due to the increase of the effective hopping amplitude between magnetic ions as the interatomic distance contracts. In contrast, the anisotropic IEI exhibit a much weaker dependence on pressure. Most of them remain essentially unchanged up to 15.0 GPa. The most notable deviations are observed for the $3^{\text{rd}}$ NN DM interaction, which increases by more than 40 $\mu$eV, and the symmetric anisotropic exchange interaction for the first two NN shells, which undergoes variations of the order of 1 $\mu$eV in the considered pressure range.

\clearpage

\begin{figure}[h]
    \centering
    
    \begin{subfigure}{0.7\linewidth}
        \centering
        \includegraphics[width=\linewidth]{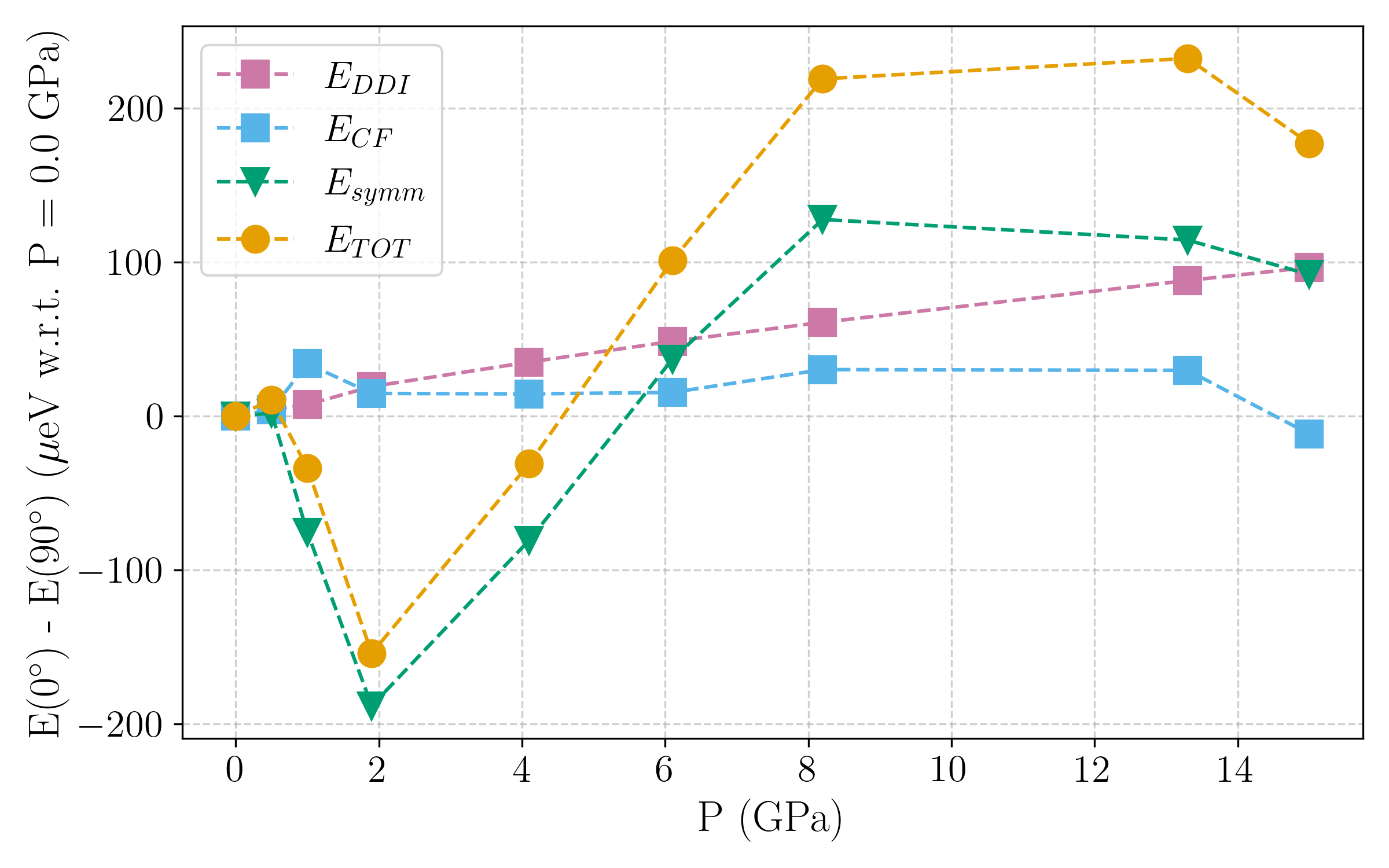}
    \end{subfigure}

    \caption{Dependence of the total magnetic anisotropy energy (orange) and its contributions from the classical DDI (pink), CF interaction (light blue) and symmetric anisotropic exchange (green) on pressure, shown relative to their respective values at P = 0.0 GPa.}
    \label{fig:sm_mae_vs_p}
\end{figure}

Next, we investigate the pressure 
response of the total magnetic anisotropy energy and its contributions,  arising from classical DDI, CF and symmetric anisotropic exchange interactions. As shown in Suppl. Fig.~\ref{fig:sm_mae_vs_p}, the classical DDI provides a weak positive contribution that scales with pressure, reaching 0.1 meV at its maximum. A similar behaviour is also observed for the CF contribution, reporting variations of the order of 10 $\mu$eV, while still remaining below 50 $\mu$eV. In contrast, the symmetric anisotropic exchange contribution displays a non-monotonic pressure dependence. At low pressures ($P < 2.0$ GPa), its pressure response is negative 
reaching a minimum of approximately -0.2 meV at P $\simeq$ 2.0 GPa. Beyond this pressure, the response exhibits a similarly rapid increase that eventually saturates at roughly 0.1 meV for $P > 8$ GPa. Despite these variations, it is worth emphasizing that no qualitative change in the magnetic anisotropy is observed throughout the considered pressure range. In particular, the negative peak due to the symmetric anisotropic exchange contribution is not sufficient to reverse the sign of the total magnetic anisotropy energy, which remains positive thanks to the dominant term arising from classical DDI.

\begin{figure}[h]
    \centering
    \includegraphics[width=0.7\linewidth]{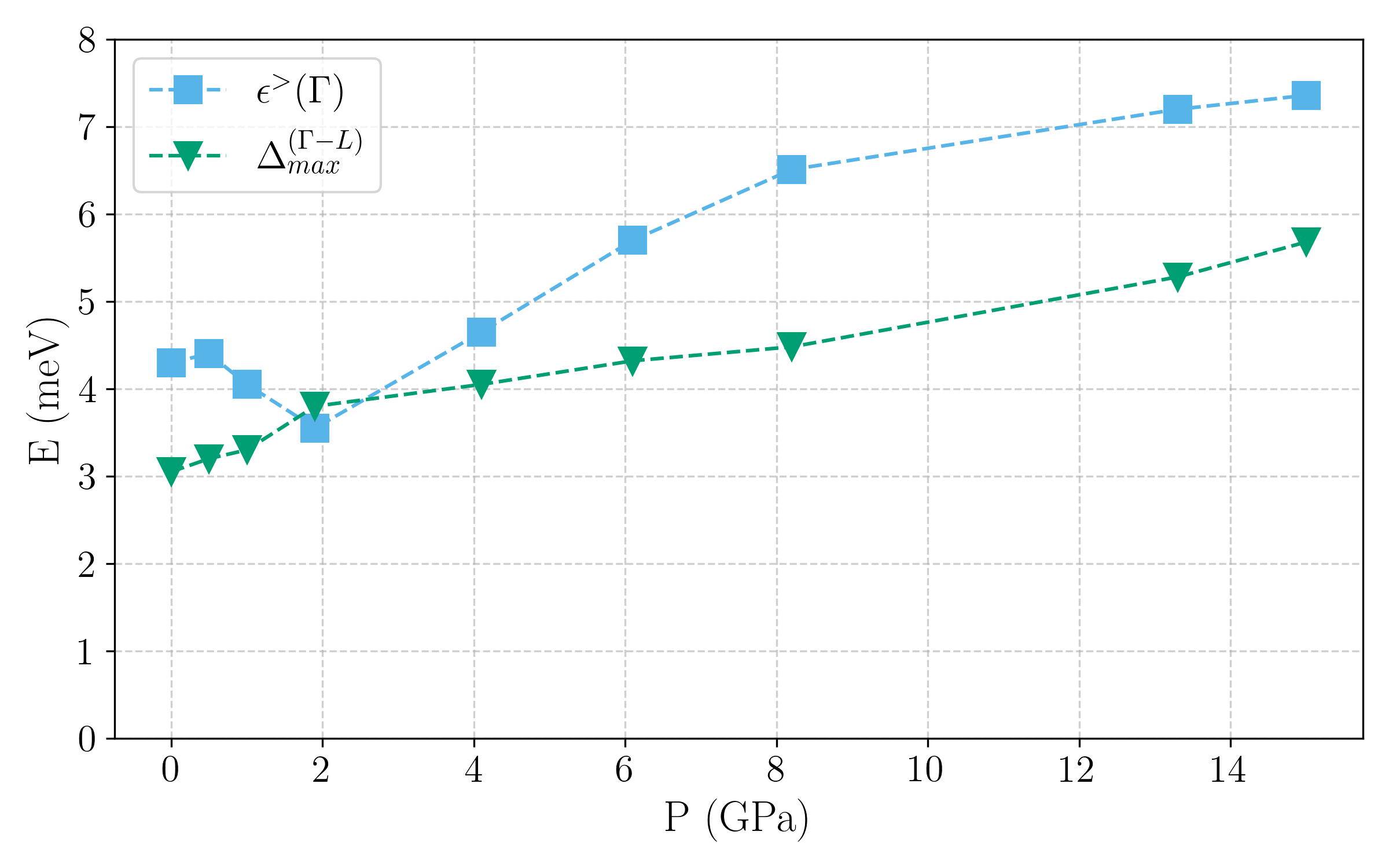}
    \caption{Energy of the upper magnon branch $\epsilon^{>}$ at $\Gamma$ and the maximum magnon splitting $\Delta^{(\Gamma-L)}_{max}$ at large wave vectors along the $\Gamma$–L segment as a function of pressure.}
    \label{fig:sm_magnsplits_vs_p}
\end{figure}

In analogy to the symmetric anisotropic exchange contribution to the magnetic anisotropy energy, we find that the upper magnon-branch energy $\epsilon^{>}$ at $\Gamma$ also exhibits a non-monotonic pressure dependence. Below P = 2.0 GPa, $\epsilon^{>}(\Gamma)$ decreases weakly with pressure, but it then increases at higher pressures. At the same time, the maximum magnon splitting $\Delta^{(\Gamma-L)}_{max}$ at large wave vectors along $\Gamma$-L exhibits a weak increase with pressure, consistently with the enhancement of the difference $\Delta J_{10}$ between the two inequivalent $10^{\text{th}}$ NN isotropic exchange interactions (see Suppl. Fig.~\ref{fig:sm_ieis_vs_p}b). The pressure effects on these magnon features of $\alpha$-MnTe are shown in Suppl. Fig.~\ref{fig:sm_magnsplits_vs_p}.

\vspace{5cm}

\bibliographystyle{apsrev4-2-pagesdoi}
\bibliography{supplementary}